\documentclass[a4paper,10pt]{article}
\usepackage{amsfonts, amsthm, amsmath}
\newcommand{\field}[1]{\mathbb{#1}}
\numberwithin{equation}{section}
\title{ {\bf
Superintegrable systems with spin and second-order 
integrals of motion }}

\author{\vspace{1cm}\\
         {\bf Jean-Francois D\'{e}silets}$^{1}$
        \thanks{E-mail address:
        desilets@crm.umontreal.ca} \,,
         {\bf Pavel Winternitz}$^{1,2}$
        \thanks{E-mail address:
        wintern@crm.umontreal.ca}
         {\,, and \bf \.{I}smet Yurdu\c{s}en}$^3$
        \thanks{E-mail address:
       yurdusen@hacettepe.edu.tr}
\\
\\$^1$Centre de Recherches Math\'{e}matiques, 
Universit\'{e} de Montr\'{e}al,\\ CP 6128, Succ. Centre-Ville, Montr\'{e}al, 
Quebec H3C 3J7, Canada
\\
\\$^2$D\'{e}partement de Math\'{e}matiques et de Statistique, 
Universit\'{e} de Montr\'{e}al,\\ CP 6128, Succ. Centre-Ville, Montr\'{e}al, 
Quebec H3C 3J7, Canada
\\
   \\$^3$Department of Mathematics, Hacettepe University,
                     \\ 06800 Beytepe, Ankara, Turkey}

\date{\today}
\newpage
\begin{document}
\setlength{\baselineskip}{24pt} 
\maketitle
\setlength{\baselineskip}{7mm}
\begin{abstract}
We investigate a quantum nonrelativistic system describing the interaction of two particles with spin $\frac{1}{2}$ and spin 0, respectively. We assume that the Hamiltonian is rotationally invariant and parity conserving and identify all such systems which allow additional integrals of motion that are second order matrix polynomials in the momenta. These integrals are assumed to be scalars, pseudoscalars, vectors or axial vectors. Among the superintegrable systems obtained, we mention a deformation of the Coulomb potential with scalar potential $V_0=\frac{\alpha}{r}+\frac{3\hbar^2}{8r^2}$ and spin orbital one $V_1=\frac{\hbar}{2r^2}$.
\end{abstract}


\newpage
\section{Introduction}
\label{intro}
This article is part of a research program, the purpose of which is to study nonrelativistic integrable and superintegrable systems with spin in real three-dimensional Euclidean space. A recent article \cite{wy3} was devoted to a system of two nonrelativistic particles with spin $s=\frac{1}{2}$ and $s=0$, respectively. Physically, this can be interpreted \emph{e.g.} as a nucleon-pion interaction or an electron - $\alpha$ particle one. An earlier article \cite{Winternitz.c} was devoted to the same problem in two dimensions.

We recall that a Hamiltonian system with $n$ degrees of freedom is called integrable if it allows $n$ independent commuting integrals of motion (including the Hamiltonian) and superintegrable if more then $n$ independent integrals exist. 

In this paper, we will consider the Hamiltonian
\begin{equation}
H=-\frac{\hbar^2}{2} \Delta + V_0(r) + V_1(r)\, (\vec{\sigma}, \vec{L})\,,
\label{intro1}
\end{equation}
in real three-dimensional Euclidean space $E_3$. Here $H$ is a matrix operator acting on a 
two-component spinor and we decompose it in terms of the $2 \times 2$ identity matrix $I$
and Pauli matrices (we drop the matrix $I$ whenever this does not cause confusion). We assume that the scalar potential $V_0(r)$ and the spin-orbital one $V_1(r)$ depend on the (scalar) distance $r$ only. The same system was already considered in \cite{wy3} and the search for superintegrable systems was restricted to those that allow integrals that are first-order polynomials in the momenta.

Here we will concentrate on the case when the integrals of motion are allowed to be second-order matrix polynomials in the momenta. Our notations are the same as in \cite{wy3}, \emph{i.e.}
\begin{equation}
 p_k = -i\hbar \partial_{x_k}\,, \qquad L_k = -i\hbar \epsilon_{klm}x_l\partial_{x_m}\,,
\end{equation}
are the linear and angular momentum respectively and
\begin{equation}
\sigma_1 = 
\left( \begin{array}{cc}
0 & 1 \\
1 & 0 \end{array} \right)\,, \qquad
\sigma_2 = 
\left( \begin{array}{cc}
0 & -i \\
i & 0 \end{array} \right)\,, \qquad
\sigma_3 = 
\left( \begin{array}{cc}
1 & 0 \\
0 & -1 \end{array} \right)\,,
\end{equation}
are the Pauli matrices.

In the case of a purely scalar potential
\begin{equation}
H=\frac{1}{2}\vec{p}^{\,2} + V_0(\vec{x})\,,
\label{introscalar}
\end{equation}
the same assumption, namely that the potential is spherically symmetric, $V_0=V(r)$ would lead only to two superintegrable systems, namely the Kepler-Coulomb one $V_0=\frac{\alpha}{r}$ and the harmonic oscillator, $V_0 = \omega r^2$. Indeed, Bertrand's theorem \cite{Bertrand,Goldstein} tells us that the only two spherically symmetric potentials in which all bounded trajectories are closed are precisely these two. On the other hand, if a classical Hamiltonian (\ref{introscalar}) is maximally superintegrable ($2n-1$ functionally independent integrals of motion in $E_n$ that are well defined functions on phase space) then all bounded trajectories must be closed \cite{Nekhoroshev}.

The superintegrability of the hydrogen atom in quantum mechanics is due to the existence of the Laplace-Runge-Lenz integral of motion. This was implicitly used by Pauli \cite{Pauli} and explicitly by Fock \cite{Fock} and Bargmann \cite{Bargmann} to calculate the energy levels and wave functions of the hydrogen atom. Similarly, the superintegrability of the isotropic harmonic oscillator is due to the existence of the quadrupole tensor \cite{Jauch} (also known as the Fradkin tensor \cite{Fradkin}).

If the potential $V_0(\vec{x})$ in (\ref{introscalar}) is not spherically symmetric, many new possibilities occur. The first one studied was the anisotropic harmonic oscillator with rational ratio of frequencies \cite{Jauch}.

A systematic study of superintegrable systems in $E_n$ with $n=2,3$ and $n$ general was started more then 40 years ago \cite{Fris}-\cite{Evans.b}. Most of the earlier work was on second-order superintegrability, \emph{i.e.} with integrals of motion that are second-order polynomials in the momenta. The existence of complete sets of commuting second-order integrals of motion is directly related to the separation of variables in the Hamilton-Jacobi or Schr\"odinger equation, respectively \cite{Fris}-\cite{Kalnins.a}. Superintegrability of the system (\ref{introscalar}) is thus related to multiseparability.

Second-order superintegrability has been studied in 2- and 3-dimensional spaces of constant and nonconstant curvature \cite{Fris}-\cite{Grosche3},\cite{Miller.b}-\cite{Tempesta} and also in $n$ dimensions \cite{Rodriguez,Kalnins.h,Kalnins.i}. For third-order superintegrability see \cite{Drach}-\cite{Marquette.a}.

Recently, infinite families of classical and quantum systems with integrals of arbitrary order have been discovered, shown to be superintegrable and solved \cite{Tremblay.b}-\cite{Marquette.c}.

Previous studies of superintegrable systems with spin have been of three types. One describes a particle with spin interacting with an external field, \emph{e.g.} an electromagnetic one \cite{Pronko.a}-\cite{Nikitin.e}. The second describes a spin $\frac{1}{2}$ particle interacting with a dyon \cite{DHoker} or with a self-dual monopoles \cite{Feher}. The third one is our program to study superintegrability in a system of two particles of which at least one has nonzero spin \cite{wy3,Winternitz.c}.

In this paper, we look for superintegrable systems of the form (\ref{intro1}). The system is integrable by construction. Since $V_0(r)$ and $V_1(r)$ depend on the distance $r$ alone and $(\vec{\sigma},\vec{L})$ is a scalar, the Hamiltonian $H$ commutes with the total angular momentum $\vec{J}$. As a matter of fact, {\it trivial} integrals (present for arbitrary functions $V_0(r)$ and $V_1(r)$) are
\begin{equation}
 H\,, \quad \vec{J}=\vec{L}+\frac{\hbar}{2} \vec{\sigma}\,, \quad (\vec{\sigma},\vec{L}) \,, \quad \vec{L}^2 \,.
\label{trivialscalars}
\end{equation}
For some potentials $V_0(r)$ and $V_1(r)$ integrals of order $0$ or $1$ in the
momenta exist \cite{wy3}. They can be multiplied by one of the trivial integrals
(\ref{trivialscalars}) (or by some function of them). This will provide further integrals
which are nontrivial, but obvious. We will mention them whenever they
occur and they may be useful for solving the corresponding superintegrable
systems. 

In view of the rotational and parity invariance of the Hamiltonian (\ref{intro1}), we shall search for integrals of motion that have a well defined behavior under these transformations. Thus, we will separately look for systems that allow integrals that are scalars, pseudoscalars, vectors or axial vectors. Tensor integrals are left for a future study.

To obtain nontrivial results, we impose from the beginning that the spin-orbital interaction be present ($V_1 \neq 0$). We also recall a result from \cite{wy3}, namely for 
\begin{equation}
V_1(r) = \frac{\hbar}{r^2}\,, \qquad V_0(r) \quad \text{arbitrary}\,,
\end{equation}
the Hamiltonian (\ref{intro1}) allows 2 first-order axial vector integrals of motion. They are $\vec{J}$ and
\begin{equation}
\label{Saxial}
 \vec{S}=-\frac{\hbar}{2} \vec{\sigma}+\hbar \frac{\vec{x}}{r^2} (\vec{x},\vec{\sigma})\,.
\end{equation}
For
\begin{equation}
 V_1(r) = \frac{\hbar}{r^2}\,, \qquad V_0(r) =\frac{\hbar^2}{r^2}\,,
\end{equation}
it allows 2 first-order axial vector integrals and 1 first-order vector integral. They are $\vec{J}$, $\vec{S}$ and 
\begin{equation}
\label{Piaxial}
 \vec{\Pi} = \vec{p}-\frac{\hbar}{r^2} (\vec{x}\wedge\vec{\sigma})\,.
\end{equation}

These two systems are first-order superintegrable and the term $V_1(r) = \frac{\hbar}{r^2}$ can be induced from a Hamiltonian with $V_1(r) = 0$ by a  gauge transformation \cite{wy3}.

%%%%%%%%%%%%%%%%%%%%%%%%%%%%%%%%%%%%%%%%%%%%%%%%%%%%%%%%%%%%%%%%%%%%%%%%%%%%%%%%%%%%%%%%%%%%%%%%%%%%%%%%%%%%%%%%%%%%%%%%%%%%%%%%%%%%%%%%%%%
\section{$O(3)$ multiplets of integrals of motion} 
\label{o3multiplets}
Rotations and reflections in $E(3)$ 
leave the Hamiltonian (\ref{intro1}) invariant but can 
transform the integrals of motion into new invariants. Thus 
instead of solving the whole set of determining equations we 
simplify the problem by classifying the integrals of motion 
into irreducible $O(3)$ multiplets. 

At our disposal are two vectors $\vec{x}$ and $\vec{p}$ and one 
pseudovector $\vec{\sigma}$. The integrals we are considering 
can involve at most second-order powers of $\vec{p}$ and first-order 
powers of $\vec{\sigma}$, but arbitrary powers of $\vec{x}$.

We shall construct scalars, pseudo-scalars, vectors and 
axial vectors in the space
\begin{eqnarray}
\Big\{\{\vec{x}\}^n \times \vec{p} \times \vec{\sigma}\Big\}\,.
\label{generalspace}
\end{eqnarray}

The quantities $\vec{x}$, $\vec{p}$ and $\vec{\sigma}$ allow us to 
define $6$ independent ``directions'' in the direct product of the 
Euclidean space and the spin one, namely 
\begin{eqnarray}
\Big\{\vec{x},\, \vec{p},\, \vec{L}=\vec{x}\wedge\vec{p},\, \vec{\sigma},\, \vec{\sigma}\wedge\vec{x},\, \vec{\sigma}\wedge\vec{p}\Big\}\,,
\label{sixdimspace}
\end{eqnarray}
and any $O(3)$ tensor can be expressed in terms of these. The 
positive integer $n$ in (\ref{generalspace}) is arbitrary and 
any scalar in $\vec{x}$ space will be written as $f(r)$ where $f$ 
is an arbitrary function of $r=\sqrt{x^2+y^2+z^2}$. Since 
$\vec{p}$ figures at most quadratically and $\vec{\sigma}$ at most 
linearly we can form exactly seven linearly independent scalars and seven 
pseudoscalars out of the quantities (\ref{sixdimspace}): 

Scalars:
\begin{eqnarray}
&S_1=1\,, \qquad S_2=\vec{p}^{\,\,2}\,, \qquad S_3=(\vec{x}, \vec{p})\,, \qquad S_4=(\vec{\sigma}, \vec{L})\,, \nonumber \\
&S_5=(\vec{x}, \vec{p})\,(\vec{\sigma}, \vec{L})\,, \qquad S_6=\vec{L}^{\,2}\,, \qquad S_7=(\vec{x}, \vec{p})^2\,.
\label{scalars}
\end{eqnarray}

Pseudoscalars:
\begin{eqnarray}
&P_1=(\vec{\sigma}, \vec{p})\,, \qquad P_2=(\vec{\sigma}, \vec{x})\,,
\qquad P_3=\vec{p}^{\,2}\,(\vec{x}, \vec{\sigma})\,, \qquad P_4=(\vec{x}, 
\vec{p})\,(\vec{x}, \vec{\sigma})\,,  \nonumber \\
&P_5=(\vec{x}, \vec{p})^2\,(\vec{x}, \vec{\sigma})\,, \qquad P_6=(\vec{x}, \vec{p})\,(\vec{p}, \vec{\sigma})\,,\qquad 
P_7=(\vec{x}, \vec{\sigma})\,\vec{L}^{\,2}\,.
\label{pseudoscalars}
\end{eqnarray}

The independent vectors and axial vectors are as follows.

Vectors:
\begin{eqnarray}
&&\vec{V}_1 = \vec{x}\,, \qquad
\vec{V}_2 = \vec{p}\,, \qquad
\vec{V}_3 = \vec{x} \wedge \vec{\sigma}\,, \qquad
\vec{V}_4 = \vec{p} \wedge \vec{\sigma}\,, \qquad
\vec{V}_5 = \vec{p}^{\,2}\,\vec{x}\,, \nonumber \\
&&\vec{V}_6 = \vec{p}^{\,2}\,(\vec{x} \wedge \vec{\sigma})\,, \qquad
\vec{V}_7 = (\vec{x}, \vec{p})\, \vec{x}\,, \qquad
\vec{V}_8 = (\vec{x}, \vec{p})\, \vec{p}\,, \qquad
\vec{V}_9 = (\vec{x}, \vec{p})\, (\vec{x} \wedge \vec{\sigma})\,, \nonumber \\
&&\vec{V}_{10} = (\vec{x}, \vec{p})\, (\vec{p} \wedge \vec{\sigma})\,, \qquad
\vec{V}_{11} = (\vec{\sigma}, \vec{L})\, \vec{x}\,, \qquad
\vec{V}_{12} = (\vec{\sigma}, \vec{L})\, \vec{p}\,, \qquad
\vec{V}_{13} = (\vec{x}, \vec{p})^2\, \vec{x}\,, \nonumber \\
&&\vec{V}_{14} = (\vec{x}, \vec{p})^2\, (\vec{x} \wedge \vec{\sigma})\,, \qquad
\vec{V}_{15} = (\vec{x}, \vec{p})\, (\vec{\sigma}, \vec{L})\, \vec{x}\,, \qquad
\vec{V}_{16} = \vec{L}^{\,2}\, \vec{x}\,, \qquad
\vec{V}_{17} = \vec{L}^{\,2}\, (\vec{x} \wedge \vec{\sigma})\,, \nonumber \\
&&\vec{V}_{18} = (\vec{\sigma}, \vec{x})\, \vec{L}\,, \qquad
\vec{V}_{19} = (\vec{x}, \vec{p})\, (\vec{x}, \vec{\sigma})\, \vec{L}\,.
\label{vectors}
\end{eqnarray}

Axial vectors:
\begin{eqnarray}
&&\vec{A}_1=\vec{\sigma}\,, \qquad
\vec{A}_2=\vec{L}\,, \qquad
\vec{A}_3=\vec{p}^{\,2}\,\vec{\sigma}\,, \qquad
\vec{A}_4=(\vec{x}, \vec{p})\, \vec{\sigma}\,, \qquad
\vec{A}_5=(\vec{x}, \vec{p})\, \vec{L}\,, \qquad
\vec{A}_6=(\vec{\sigma}, \vec{L})\, \vec{L}\,, \qquad \nonumber \\
&&\vec{A}_7=(\vec{x}, \vec{p})^2\, \vec{\sigma}\,, \qquad
\vec{A}_8=(\vec{\sigma}, \vec{p})\, \vec{x}\,, \qquad
\vec{A}_9=(\vec{\sigma}, \vec{p})\, \vec{p}\,, \qquad
\vec{A}_{10}=(\vec{\sigma}, \vec{x})\, \vec{x}\,, \qquad
\vec{A}_{11}= (\vec{\sigma}, \vec{x})\, \vec{p}\,, \nonumber \\
&&\vec{A}_{12}=\vec{p}^{\,2}\, (\vec{x}, \vec{\sigma})\, \vec{x}\,, \qquad
\vec{A}_{13}=(\vec{x}, \vec{p})\, (\vec{\sigma}, \vec{x})\, \vec{x}\,, \qquad
\vec{A}_{14}=(\vec{x}, \vec{p})^2\, (\vec{x}, \vec{\sigma})\, \vec{x}\,, \qquad
\vec{A}_{15}=(\vec{x}, \vec{p})\, (\vec{p}, \vec{\sigma})\, \vec{x}\,, \nonumber \\
&&\vec{A}_{16}=\vec{L}^{\,2}\, \vec{\sigma}\,,\qquad
\vec{A}_{17}=(\vec{x}, \vec{p})\, (\vec{x}, \vec{\sigma})\, \vec{p}\,, \qquad 
\vec{A}_{18}=(\vec{x}, \vec{\sigma})\, \vec{L}^{\,2}\, \vec{x}\,. 
\label{axialvectors}
\end{eqnarray}

An arbitrary function $f(r)$ is also a scalar and each of the 
quantities in (\ref{scalars})-(\ref{axialvectors}) can be 
multiplied by $f(r)$ without changing its properties under 
rotations or reflections.

Even though all the above expressions are linearly independent, higher order polynomial relations between them exist. More importantly, we shall use linear relations with coefficients depending on the distance $r$ that exist, namely
\begin{eqnarray} 
S_7 = r^2S_2-S_6\,, \qquad 
P_7 = r^2P_3-P_5\,, \qquad
V_{16} = r^2V_5-V_{13}\,, \qquad
V_{17} = r^2V_6-V_{14}\,, \qquad \nonumber \\
V_{18} = -r^2V_4+V_9+V_{11}\,, \qquad
V_{19} = -r^2V_{10}+V_{14}+V_{15}\,, \qquad
A_{16} = r^2A_3-A_7\,, \qquad \nonumber \\
A_{17} = r^2(A_9-A_3)+A_6+A_7+A_{12}-A_{15}\,, \qquad
A_{18} = r^2A_{12}-A_{14}\,.
\label{syzygies}
\end{eqnarray}

We mention that the vector $(\vec{\sigma},\vec{p})\vec{L}$ was eliminated from the list (\ref{vectors}) using the nontrivial linear relation $(\vec{\sigma},\vec{p})\vec{L} = V_6-V_{10}+V_{12}$. We use the relations (\ref{syzygies}) to remove the left hand sides of (\ref{syzygies}) from the analysis completely.

The relations determining $S_7,P_7,V_{16},V_{17},A_{16}$ and $A_{18}$ are all consequences of the simple vector relation $\vec{L}^2 = \vec{x}\,^2\,\vec{p}\,^2-(\vec{x},\vec{p})^2$. Those determining $V_{18}$ and $V_{19}$ follow from the identity $(\vec{\sigma},\vec{x}) (\vec{x} \wedge \vec{p}) = -\vec{x}\,^2 (\vec{p}\wedge \vec{\sigma})+(\vec{x},\vec{p})( \vec{x}\wedge\vec{\sigma})+(\vec{\sigma},\vec{x}\wedge\vec{p})\vec{x}$.

%%%%%%%%%%%%%%%%%%%%%%%%%%%%%%%%%%%%%%%%%%%%%%%%%%%%%%%%%%%%%%%%%%%%%%%%%%%%%%%%%%%%%%%%%%%%%%%%%%%%%%%%%%%%%%%%%%%%%%%%%%%%%%%%%%%%%%%%%%%

\section{Symmetrization of the integrals of motion}
\label{symmetsec}
In the rest of the article we separately take the linear combinations 
of all the scalars, pseudo-scalars, vectors and axial vectors with coefficients $f_i(r)$ that are real functions of $r$. However, 
instead of having the bare linear combinations of these tensors, the full 
symmetric forms are needed for the analysis of the commutation relations. 
Thus, in this section, we briefly describe how this symmetrization process 
carried out basically by working on the linear combination of the scalars chosen 
as a prototype.

In writing the most general scalar operator, one needs to symmetrize 
the linear combination of the scalars given in (\ref{scalars}) 
\begin{eqnarray}
X_S = \sum_{j=1}^{6}{f_j(r)S_j}\,, \quad f_j(r) \in \field{R}\,,
\label{firstoption}
\end{eqnarray}
term by term. It is obvious that the terms, for example, $x_i\,p_i$ 
(here and throughout the whole article summation over the repeated indices 
through $1$ to $3$ is to be understood) and $p_i\,x_i$ are in fact different 
scalars and their symmetric form is $\frac{1}{2}(x_i\,p_i+p_i\,x_i)$. However, 
at this stage an immediate question arises: should we associate a single 
arbitrary function of $r$ multiplying each term obtained from symmetrization 
or different weightings are necessary for each of them.   

Let us consider the scalar $S_3$, for example. The most general 
possible form of it is achieved by giving an arbitrary function 
of $r$ to each permutation
\begin{eqnarray}
S_3 = f_a(r)x_ip_i+f_b(r)p_ix_i+x_if_c(r)p_i+x_ip_if_d(r)+p_if_e(r)x_i+p_ix_if_f(r)\,.
\end{eqnarray}
Requiring that the operator be hermitian, that is $S_3 = S_3^\dagger$, we get
\begin{eqnarray}
f_a(r)x_ip_i+f_b(r)p_ix_i+x_if_c(r)p_i+x_ip_if_d(r)+p_if_e(r)x_i+p_ix_if_f(r) \nonumber \\
=p_ix_if_a(r)+x_ip_if_b(r)+p_if_c(r)x_i+f_d(r)p_ix_i+x_if_e(r)p_i+f_f(r)x_ip_i\,.
\label{s3fullform}
\end{eqnarray}
Hence self-adjointness reduces the half of the arbitrary functions
\begin{eqnarray}
f_a(r)=f_f(r)\,, \qquad f_b(r)=f_d(r)\,, \qquad f_c(r)=f_e(r)\,.
\end{eqnarray}
Thus, $S_3$ can now be rewritten as  
\begin{eqnarray}
S_3=(f_a(r)+f_c(r))x_ip_i+p_ix_i(f_a(r)+f_c(r))+f_b(r)p_ix_i+x_ip_if_b(r)\,.
\label{hermitiancouples}
\end{eqnarray}
Furthermore it is always possible to move the derivative terms to the right
\begin{eqnarray}
S_3=2(f_a(r)+f_b(r)+f_c(r))x_ip_i -3\hbar i(f_a(r)+f_b(r)+f_c(r)) \nonumber \\ -i\hbar r(f_a'(r)+f_b'(r)+f_c'(r))\,.
\label{ptotheright}
\end{eqnarray}
Defining
\begin{eqnarray}
f_3(r) = f_a(r)+f_b(r)+f_c(r)
\label{redefine}
\end{eqnarray}
we write
\begin{eqnarray}
S_3=2f_3(r)x_ip_i -3i\hbar f_3(r)-i\hbar rf_3'(r)\,.
\end{eqnarray}

Thus for the scalar $S_3$ it is explicitly shown that 
a single arbitrary function of $r$ is enough to make the 
symmetric form as general as can be.  

However, it is easy to see how difficult this process would become 
even for the scalars such as $S_5$, for which we would have to start 
by writing $5!=120$ arbitrary functions of $r$. We have already shown that 
a single arbitrary function of $r$ is enough to give the symmetric 
form of the operators that are first-order in the momentum \cite{wy3}. 

To investigate the operators that are second-order in the momentum, 
we wrote a computer program code working under Mathematica. The code 
organizes the permutations of a given operator into Hermitian couples, 
which does the equivalent steps given in the previous example 
(\ref{s3fullform}-\ref{hermitiancouples}). Hence for an operator made of n terms, 
we are left with $\frac{n!}{2}$ Hermitian couples. Moving the derivative terms 
to the extreme right for each couple, we see that all of the Hermitian couples for 
a given operator have the same form (as in (\ref{ptotheright})) although 
some of them give a few extra terms. These extra terms can always be absorbed into 
already existing terms (of lower order in $\vec{p}$). Hence a redefinition 
of the arbitrary functions associated with a 
given operator (as done in (\ref{redefine})) assures that it is enough to have 
only one arbitrary function of $r$ for each scalar.

For example, let us consider the symmetrization of the scalar $S_2$. 
There are two distinct Hermitian couples associated with $S_2$, namely:
\begin{eqnarray}
 p_if_2(r)p_i + p_if_2(r)p_i\\
p_ip_if_2(r)+f_2(r)p_ip_i\,.
\end{eqnarray}
Moving the derivatives to the right in the first couple, we obtain: 
\begin{equation}
   p_if_2(r)p_i + p_if_2(r)p_i = 2(f_2(r)p_i^2-i\hbar f_2'(r)\frac{x_i}{r}p_i)\,.
\end{equation}
The same process with the second couple gives: 
\begin{equation}
 p_ip_if_2(r)+f_2(r)p_ip_i = 2(f_2(r)p_i^2-i\hbar f_2'(r)\frac{x_i}{r}p_i) -\hbar^2 f_2''(r)\,. 
\end{equation}

Here, we have an extra term in the second derivative of $f_2(r)$. 
However, this is just another arbitrary function of $r$ and it is already associated 
with the scalar $S_1$. Hence we could eliminate it by redefining the function 
associated with $S_1$ as; $f_1(r) = f_1(r) -\hbar^2 f_2''(r)$ and the sum 
(\ref{firstoption}) would remain unchanged. The Mathematica computer program code 
makes it clear that the extra terms coming from choosing one 
Hermitian couple over another for a given operator could always be eliminated 
in this way.

When writing the operators using tensor notation, 
one understands that the number of free indices is 
irrelevant to the symmetrization of a given operator, 
assuring the validity of the above method for any class of operators 
(pseudoscalar, vector and axial vector). In its final version, 
the computer program is able to take a list of 
operators and symmetrize all of them in an optimal manner keeping no 
unnecessary terms.

%%%%%%%%%%%%%%%%%%%%%%%%%%%%%%%%%%%%%%%%%%%%%%%%%%%%%%%%%%%%%%%%%%%%%%%%%%%%%%%%%%%%%%%%%%%%%%%%%%%%%%%%%%%%%%%%%%%%%%%%%%%%%%%%%%%%%%%%%%%%%%%%%%%%%%%%%%%%%%%%%%%%%%%%%%%%%%%%%%%%%%%%%%%%%%%%%%%%%%%%%%%%%%%%%%%%%%%%%%%%%%%%%%%%%%%%%%%%%%%%%%%%%%%%%%%%%%%%%%%%%%%%%%%%%%%%%%%%%%%%%%%%%%%%%%%%%%%%%%%%%%%%%%%%%%%%%%%%%%%%%%%%%%%%%%%%%%%%%%%%%%%%%%%%%%%%%%%%%%%%%%%%%%%%%%%%%%%%%%%%%%%%%%%%%%%%%%%%%%%%%%%%%%%%%%%%%

\section{Commutativity conditions for scalars and pseudoscalars} 
\label{comconditionsforsp}

\subsection{Scalars}
\label{subsecscalars}
Let us take a linear combination of the independent scalars given in (\ref{scalars})
\begin{equation}
{\widetilde X}_S=\sum_{j=1}^6 f_j(r) S_j\,, 
\label{linscalars}
\end{equation}
and fully symmetrize it as described in Section \ref{symmetsec}. 
The symmetric form of (\ref{linscalars}) is written as 
\begin{eqnarray}
X_S &=& f_1 - 3i\hbar f_3 - i\hbar r f_3^{\prime} + 2 f_2\, (\vec{p}, \vec{p}) 
+ 2 f_5\, \big(\vec{x}, (\vec{\sigma}, \vec{L})\,\vec{p} \big) + 
f_6\, (\vec{L}, \vec{L}) \nonumber \\ 
&+& \Big(f_4 - 5i\hbar f_5 - i\hbar r f_5^{\prime} \Big)\, (\vec{\sigma}, \vec{L}) + 
\Big(2 f_3 - \frac{2i\hbar}{r} f_2^{\prime}\Big)\, 
(\vec{x}, \vec{p})\,.
\label{fullsymmetricscalars}
\end{eqnarray}

The requirement that $[H, X_S]=0$, gives us the determining 
equations for this case. The determining equations, obtained by 
equating the coefficients of the third-order terms to zero in the 
commutativity equation, become 
\begin{eqnarray}
f_5 = 0\,, \qquad f_2^{\prime} = 0\,, \qquad 
f_6^{\prime} = 0\,. 
\label{thirdordetscalar}
\end{eqnarray}
The determining equations, obtained by 
equating the coefficients of the second-order and first-order 
terms to zero in the commutativity equation, read respectively  
\begin{eqnarray}
f_3 = 0\,, \qquad f_4^{\prime} = 4 f_2\, V_1^{\prime}\,, 
\label{secondordscalars}
\end{eqnarray}
and 
\begin{eqnarray}
f_1^{\prime} = 4 f_2\, V_0^{\prime}\,. 
\label{firstordscalars}
\end{eqnarray}
The rest of the determining equations are then satisfied 
identically. It is immediately seen that the only 
solutions for equations 
(\ref{thirdordetscalar}) - (\ref{firstordscalars}) are
\begin{eqnarray} 
f_1 = 4 c_1 V_0 + c_3\,, \qquad f_2 = c_1\,, \qquad f_4 = 4 c_1 V_1 + c_4\,, 
\qquad f_6 = c_2\,,
\label{solfordetscalars}
\end{eqnarray}
where $c_i$, $i = 1,\ldots, 4$ are real constants. 

Thus the corresponding four integrals of motion are those given in \ref{trivialscalars}, \emph{i.e.} there are no nontrivial scalar integrals of motion.

%%%%%%%%%%%%%%%%%%%%%%%%%%%%%%%%%%%%%%%%%%%%%%%%%%%%%%%%%%%%%%%%%%%%%%%%%%%%%%%%%%%%%%%%%%%%%%%%%%%%%%%%%%%%%%%%%%%%%
%%%%%%%%%%%%%%%%%%%%%%%%%%%%%%%%%%%%%%%%%%%%%%%%%%%%%%%%%%%%%%%%%%%%%%%%%%%%%%%%%%%%%%%%%%%%%%%%%%%%%%%%%%%%%%%%%%%%%
%%%%%%%%%%%%%%%%%%%%%%%%%%%%%%%%%%%%%%%%%%%%%%%%%%%%%%%%%%%%%%%%%%%%%%%%%%%%%%%%%%%%%%%%%%%%%%%%%%%%%%%%%%%%%%%%%%%%%
\subsection{Pseudoscalars}
\label{subsecpseudoscalars}
As an integral of motion we take a linear combination of 
the independent pseudoscalars given in (\ref{pseudoscalars})
\begin{equation}
{\widetilde X}_P=\sum_{j=1}^6 f_j(r) P_j\,,  
\label{linpseudoscalars}
\end{equation}
which has the following fully symmetric form 
\begin{eqnarray}
X_{P} &=& (\vec{\sigma}, \vec{x})\, \Big(f_2 -i\hbar ( \frac{1}{r} f_1^{\prime} +4 f_4 +r f_4^{\prime}) \Big) 
+ \Big( 2f_1 -i\hbar\big(2f_3+4f_6 + r f_6^{\prime} \big) \Big) (\vec{\sigma}, \vec{p}) \nonumber \\
& +& (\vec{\sigma}, \vec{x})\, \Big(2f_4 -i\hbar( \frac{2}{r} f_3^{\prime} + 10f_5 + 2 rf_5^{\prime} + \frac{1}{r} f_6^{\prime}) \Big) (\vec{x}, \vec{p}) + 2f_6 \,(\vec{x}, \vec{p})\,(\vec{\sigma}, \vec{p}) \nonumber \\
& +& (\vec{\sigma}, \vec{x})\, \Big(2f_3\, (\vec{p}, \vec{p}) + 2f_5 \big(\vec{x}, (\vec{x}, \vec{p})\, \vec{p}\big)\, \Big)\,.
\label{fullsymmetricpseudoscalars}
\end{eqnarray}

Requiring that the commutator $[H, X_P]=0$, we obtain the 
determining equations for this case. Those, obtained by 
equating the coefficients of the third-order terms to zero in 
the commutativity equation, read 
\begin{eqnarray}
&&2r f_5\,V_1 + \hbar f_5^{\prime} = 0\,, \label{thirdten1} \\
&&\hbar f_3^{\prime} + 2 r\, \Big(\hbar f_5 + \big(f_3 +f_6\big)\, V_1\Big) = 0\,, \label{thirdten2} \\
&&\big(\hbar r + 4r^3 V_1 \big)\, f_5 + \hbar f_6^{\prime} - r\, \Big(2f_6V_1 - 3\hbar r f_5^{\prime}\Big) = 0\,, \label{thirdten3} \\
&&\big(4r^2 V_1 -\hbar \big)\, r\hbar f_5 - \hbar f_3^{\prime} + \hbar f_6^{\prime} - r\,\Big(2\big(f_3 + 2f_6 \big)\, V_1 - 3\hbar rf_5^{\prime}\Big) = 0\,, \label{thirdten4} \\
&&\hbar f_3 + \hbar f_6 + r\, \Big(\hbar f_3^{\prime}  + 2 r\, \big(\hbar f_5 + f_6 V_1\big) \Big) = 0\,, \label{thirdten5} \\
&&\hbar f_3^{\prime} + 2\hbar f_6^{\prime} + r\, \Big(2\big(f_3 - f_6\big)\,V_1 + 2f_5\, \big(2\hbar +r^2V_1\big) + 
3\hbar r f_5^{\prime} \Big) = 0\,, \label{thirdten6} \\
&&\hbar f_3^{\prime} + \hbar f_6^{\prime} + r\, \Big(2f_3 V_1 + f_5\, \big(3\hbar +2r^2V_1\big) + 
2\hbar r f_5^{\prime} \Big) = 0\,, \label{thirdten7} \\
&&f_3 + f_6 + r\, \Big( 3rf_5 + f_3^{\prime} + r^2 f_5^{\prime} + f_6^{\prime} \Big) = 0\,, \label{thirdten8} \\
&&\hbar f_6 + f_3\,\big(\hbar - 2r^2\, V_1\big) = 0\,, \label{thirdten9} \\
&&\Big(2r^2V_1 -\hbar \Big)\,\Big(f_3 + r^2 f_5 + f_6 \Big) - r\hbar f_6^{\prime} = 0\,. \label{thirdten10}
\end{eqnarray}

Equation (\ref{thirdten8}) can immediately be integrated to give 
\begin{eqnarray}
f_5 = \frac{c_1 - r\,(f_3 + f_6)}{r^3}\,,
\label{thirdf5}
\end{eqnarray}
where $c_1$ is a real constant. If we introduce (\ref{thirdf5}) into 
(\ref{thirdten10}) we obtain 
\begin{eqnarray}
f_6^{\prime} = c_1 \left(2 V_1 - \frac{\hbar }{r^2}\right)\,.
\label{thirdf6}
\end{eqnarray}
Then writing $f_6$ from (\ref{thirdten9}) and using the 
rest of the equations (\ref{thirdten1})-(\ref{thirdten7}), as a compatibility 
condition we obtain a nonlinear second-order equation for $V_1$  
\begin{eqnarray}
&&3\hbar^4r\big(\hbar-2r^2 V_1\big)\,V_1^{\prime\prime} + 4\hbar^2 \left(3\hbar^3 + r^2 V_1\,\Big(9\hbar^2 + 5 r^2 V_1\,\big(2 r^2 V_1 - 3\hbar\big)\Big)\right)\,V_1^{\prime} + 10\hbar^4 r^3 V_1^{\prime\,2} \nonumber \\
&& + 2 r V_1^2 \left(45\hbar^4 - 4 r^2 V_1\,\Big(15\hbar^3 + 4 r^4 V_1^2\,\big(r^2 V_1 - 3\hbar\big)\Big)\right) = 0\,.
\label{compatibilityforv1}
\end{eqnarray}

Performing a standard symmetry analysis \cite{Olver} for the equation 
(\ref{compatibilityforv1}), we find that the symmetry algebra 
is spanned by two vector fields 
\begin{eqnarray}
\vec{v}_1 = r \partial_r - 2 V_1 \partial_{V_1}\,, \qquad 
\vec{v}_2 = r^3 \partial_r +\left(\frac{\hbar}{2}-3r^2 V_1\right)\partial_{V_1}\,,
\label{vectorfields}
\end{eqnarray}
with $[\vec{v}_1, \vec{v}_2] = 2 \vec{v}_2$. It is now possible to lower the order of the equation (by two) using the standard method of symmetry reduction for ordinary differential equations \cite{Olver}. In the case of equation (\ref{compatibilityforv1}), this leads to an implicit (general) solution which we do not find useful.

An alternative is to use the symmetry algebra to find particular solutions of (\ref{compatibilityforv1}), invariant under the subgroup generated by $\vec{v}_1$, or that generated by $\vec{v}_2$. The corresponding subgroup invariants are the potentials 
\begin{eqnarray}
V_1 = \frac{C_1}{r^2}\, \qquad {\rm and}  \qquad V^*_1 = \frac{C_2}{r^3} + \frac{\hbar}{2r^2}\,,
\label{invariantss}
\end{eqnarray}
respectively. Substituting (\ref{invariantss}) into (\ref{compatibilityforv1}), we see that $V_1$ is a solution for
\begin{eqnarray}
C_1 = \Big\{-\frac{\hbar}{2}, 0, \frac{\hbar}{2}, \hbar, \frac{3 \hbar}{2} \Big\}\,.
\label{specialsol}
\end{eqnarray}

The invariant $V^*_1$ is a solution only for $C_2=0$ and that solution is already included in (\ref{specialsol}) ($C_1 = \frac{\hbar}{2}$). The solutions $V_1$ for $C_1$ as in (\ref{specialsol}) (with $C_1 \neq \frac{\hbar}{2}$) can be extended to one-parameter classes of solutions by acting on them with the symmetry group generated by $\vec{v}_2$ :
\begin{eqnarray}
\tilde{r} = \frac{r}{\sqrt{1-2\lambda r^2}}\,, \qquad 
\widetilde{V}_1 = \left(V_1-\frac{\hbar}{2r^2}\right)(1-2\lambda r^2)^{\frac{3}{2}}+\frac{\hbar}{2r^2}(1-2\lambda r^2) \,, \qquad
|\lambda| < \frac{1}{2r^2} \,.
\label{integratedgroup}
\end{eqnarray}

Substituting $V_1(r)=\frac{C_1}{r^2}$ into (\ref{integratedgroup}) and expressing $\widetilde{V}_1$ in terms of $\tilde{r}$, we obtain
\begin{eqnarray}
\widetilde{V}_1(\tilde{r}) = \frac{1}{2\tilde{r}^2} \left(\hbar+\frac{2C_1-\hbar}{\sqrt{1+2\lambda \tilde{r}^2}}\right)\,.
\end{eqnarray}

The 4 values of $C_1 \neq \frac{\hbar}{2}$ in (\ref{specialsol}) lead to 4 new potentials
\begin{eqnarray}
V_1(r) = \frac{\hbar}{2r^2} \left(1 + \frac{2\epsilon}{\sqrt{1+2\lambda r^2}}\right) \,, \qquad V_1(r) = \frac{\hbar}{2r^2} \left(1+ \frac{\epsilon}{\sqrt{1+2\lambda r^2}}\right)\,,
\label{psfamilies}
\end{eqnarray}
with $\epsilon^2=1$. Thus we have 7 potentials to consider: those in (\ref{psfamilies}) and the original $V_1 = \frac{C_1}{r^2}$ with $C_1 = -\frac{\hbar}{2}, \frac{\hbar}{2}, \frac{3\hbar}{2}$ ($C_1=0$ is trivial and $C_1 = \hbar$ is gauge induced and was considered in \cite{wy3}). Taking $\lambda=0$ in (\ref{psfamilies}) we recover the original cases with $C_1= -\frac{\hbar}{2}, \frac{3\hbar}{2}$ (and $C_1 = 0,\hbar$), but we prefer to treat them separately.

%%%%%%%%%%%%%%%%%%%%%%%%%%%%%%%%%%%%%%%%%%%%%%%%%%%%%%%%%%%%%%%%%%%%%%%%%%%%%%%%%%%%%%%%%%%%%%%%%%%%%%%%%%%%%%%%%%%%%%%%%%%%%%%%%%
{\bf Case I:} $V_1=\frac{\hbar}{2r^2}$

For this type of potential, (\ref{thirdten9}) and (\ref{thirdf5}) 
immediately imply
\begin{equation}
f_6 = 0\,, \qquad {\rm and} \qquad f_5 = \frac{c_1}{r^3} - \frac{f_3}{r^2}\,,
\end{equation}
where $c_1$ is an integration constant. Also the set 
of determining equations given in 
(\ref{thirdten1})-(\ref{thirdten10}) together with (\ref{thirdf5}) 
give us 
\begin{equation}
f_5 = - \frac{c_2}{r},
\end{equation}
where $c_2$ is an integration constant. Then the determining equations, 
obtained from lower order terms, provide us with
\begin{equation}
f_1 =  -r\, c_3\,, \qquad f_2 = \frac{4\, c_1 V_0 + c_4}{r}\,, \qquad f_4 = \frac{c_3}{r}\,,
\end{equation}
where $c_3$ and $c_4$ are integration constants. For these values of 
$f_j$ (for $j = 1, \ldots, 6$) all the determining equations, obtained from 
the requirement that the commutator $[H, X_P]=0$, are satisfied for 
any $V_0 = V_0(r)$. Since we have four arbitrary constants and none of 
them appear in the Hamiltonian, we have four different integrals 
of motion. Two of them are first-order operators and correspond to the 
ones that were found in \cite{wy3}, while the other 
two are second-order operators. They are given as, 
\begin{eqnarray}
&&X_P^1 = \frac{(\vec{\sigma}, \vec{x})}{r}\,, \label{pscase1int1} \\
&&X_P^2 = -r (\vec{\sigma}, \vec{p}) + \frac{1}{r}\,(\vec{\sigma}, \vec{x})\,(\vec{x}, \vec{p}) - \frac{i\hbar}{r}\,(\vec{\sigma}, \vec{x})\,,
\label{pscase1int2} \\ 
&&X_P^3 = 4\, \frac{(\vec{\sigma}, \vec{x})}{r} \left(\frac{1}{2}(\vec{p}, \vec{p}) + V_0\right) + \frac{2i\hbar}{r^3} (\vec{\sigma}, \vec{x})\,(\vec{x}, \vec{p}) - \frac{2i\hbar}{r} (\vec{\sigma}, \vec{p})\,, \label{pscase1int3} \\ 
&&X_P^4 = \frac{(\vec{\sigma}, \vec{x})}{r}\, \Big(6i\hbar\, (\vec{x}, \vec{p}) - 2\,\big(\vec{x}, (\vec{x}, \vec{p})\,\vec{p}\big) + 2r^2(\vec{p}, \vec{p}) \Big) - 2i\hbar r\, (\vec{\sigma}, \vec{p})\,. \label{pscase1int4}
\end{eqnarray}

However, the only really independent pseudoscalar integral is $X_P^1$ since
we have
\begin{eqnarray}
&&X_P^2 = -i X_P^1 \Big((\vec{\sigma}, \vec{L}) + \hbar \Big)\,, \label{functionalrelpscase10} \\
&&X_P^3 = 4\,X_P^1\,H\,, \label{functionalrelpscase11} \\
&&X_P^4 = 2\,X_P^1\,\Big((X_P^2)^2 - \hbar^2\Big)\,.
\label{functionalrelpscase12}
\end{eqnarray}

%%%%%%%%%%%%%%%%%%%%%%%%%%%%%%%%%%%%%%%%%%%%%%%%%%%%%%%%%%%%%%%%%%%%%%%%%%%%%%%%%%%%%%%%%%%%%%%%%%%%%%%%%%%%%%%%%%%%%%%%%%%%%%%%%%
{\bf Case II:} $V_1 = -\frac{\hbar}{2r^2}$

For this type of potential, (\ref{thirdf6}) implies 
\begin{equation}
f_6 = \frac{2\,c_1}{r} + c_2\,, 
\label{case2detthird1}
\end{equation}
where $c_1$ and $c_2$ are integration constants. Then the set 
of determining equations given in (\ref{thirdten1})-(\ref{thirdten10})
gives 
\begin{equation}
f_1 = 0\,, \qquad f_3 = - \frac{c_1}{r}\,, \qquad f_4 = 0\,, \qquad f_5 = 0\,, \qquad c_2 = 0\,. 
\label{case2detthird2}
\end{equation}
Introducing (\ref{case2detthird1}) and (\ref{case2detthird2}) into 
the determining equations obtained from lower order terms, we find the 
following equations 
\begin{eqnarray}
2 r \big(f_2 - r\,f_2^{\prime}\big) - \frac{12 \hbar^2 c_1}{r^2} = 0\,, \qquad
4c_1 r V_0^{\prime} - \frac{\hbar^2c_1 + 2c_3r^4}{r^2} = 0\,.
\label{case2detloweq}
\end{eqnarray}
Their solutions are
\begin{eqnarray}
f_2 = \frac{3\hbar^2 c_1}{2r^3} + r\,c_3\,, \qquad 
V_0 = -\frac{\hbar^2}{8\,r^2} + \alpha r^2\,, \qquad \alpha=\frac{c_3}{4c_1}\,, 
\label{case2solutions}
\end{eqnarray}
where $c_3$ is an integration constant. Hence the integral of motion 
$X_P$ depends on two constants $c_1$ and $c_3$. However, $V_0$ also 
depends on $\alpha=\frac{c_3}{4c_1}$. Thus, we can choose $c_3 = 4 \alpha c_1$ and set 
$c_1=1$. The integral of motion for this case can be written as
\begin{eqnarray}
X_P =  \frac{(\vec{\sigma}, \vec{x})}{r}\, \left(\frac{3\hbar^2}{2\,r^2} + 4\alpha\,r^2 - 2 (\vec{p},\vec{p}) \right) + 
\frac{4}{r}\,\big((\vec{x}, \vec{p}) - i\hbar\big)\, (\vec{\sigma}, \vec{p})\,.
\label{intofmotforcase2}
\end{eqnarray}

Since $X_P$ is a pseudoscalar operator it also commutes with the components of the total angular momentum $J_i$ $(i=1,2,3)$: 
$[H, X_P] = 0$, and $[J_i, X_P] = 0$, as do all the pseudoscalars obtained below.

%%%%%%%%%%%%%%%%%%%%%%%%%%%%%%%%%%%%%%%%%%%%%%%%%%%%%%%%%%%%%%%%%%%%%%%%%%%%%%%%%%%%%%%%%%%%%%%%%%%%%%%%%%%%%%%%%%%%%%%%%%%%%%%%%%
{\bf Case III:} $V_1 = \frac{3 \hbar}{2r^2}$

For this type of potential, the set 
of determining equations given in 
(\ref{thirdten1})-(\ref{thirdten10}) together with 
(\ref{thirdf6}) give us 
\begin{eqnarray}
&&f_1 = 0\,, \qquad f_3 = - \frac{c_1}{r} + \frac{c_2}{2}\,, \qquad f_4 = 0\,,  \nonumber \\
&&f_5 = \frac{4\, c_1}{r^3} - \frac{3 c_2}{2 r^2}\,, \qquad f_6 = -\frac{2 c_1}{r} + c_2\,, 
\label{case3newintofmot} 
\end{eqnarray} 
where $c_1$ and $c_2$ are integration constants. If we introduce 
these integrals of motion into the set of determining equations 
we get $c_2 = 0$. Using the determining equations, obtained from 
lower order terms, we get 
\begin{eqnarray}
f_2 = \frac{5\hbar^2\,c_1 + 2\,c_1\,r^3\,V_0^{\prime}}{r^3}\,,
\label{case3lowdet}
\end{eqnarray}
and upon introducing (\ref{case3lowdet}) back into the 
determining equations we obtain a second-order differential 
equation for $V_0$
\begin{eqnarray}
r^3\,\big(V_0^{\prime} - r\,V_0^{\prime \prime} \big) + 15\hbar^2 = 0\,.
\label{case3V0}
\end{eqnarray}
Its solution is 
\begin{eqnarray}
V_0 = \frac{15\hbar^2}{8\,r^2} + \alpha r^2\,,
\label{case3V0sol}
\end{eqnarray}
where $\alpha$ is an integration constant. Using 
(\ref{case3V0sol}) we obtain $f_2$ from (\ref{case3lowdet}) 
 \begin{eqnarray}
f_2 = c_1\,\left(4\,\alpha\,r -\frac{5\hbar^2}{2\,r^3} \right)\,.
\label{case3lowdetf2fin}
\end{eqnarray}
Since there is only one arbitrary constant for this case 
($\alpha$ appears in the Hamiltonian) we only have one 
second-order integral of motion, namely 
\begin{eqnarray}
X_P =  \frac{(\vec{\sigma}, \vec{x})}{r}\, \left(-\frac{5\hbar^2}{2\,r^2} + 4\,\alpha\,r^2 - \frac{20i\hbar}{r^2}\,(\vec{x},\vec{p}) - 2 (\vec{p},\vec{p}) + \frac{8}{r^2} \big(\vec{x}, (\vec{x}, \vec{p})\,\vec{p}\big) \right) - 
\frac{4}{r}\,\big((\vec{x}, \vec{p}) - 2i\hbar\big)\, (\vec{\sigma}, \vec{p})\,.
\label{intofmotforcase3}
\end{eqnarray}

%%%%%%%%%%%%%%%%%%%%%%%%%%%%%%%%%%%%%%%%%%%%%%%%%%%%%%%%%%%%%%%%%%%%%%%%%%%%%%%%%%%%%%%%%%%%%%%%%%%%%%%%%%%%%%%%%%%%%%%%%%%%%%%%%%
{\bf Case IV:} $V_1 = \frac{\hbar}{2r^2}\,\left(1 + \frac{\epsilon}{\sqrt{1 + \beta\,r^2}} \right)$, $\beta \equiv 2 \lambda$

For this type of potential the set 
of determining equations given in 
(\ref{thirdten1})-(\ref{thirdten10}) together with 
(\ref{thirdf6}) give us 
\begin{eqnarray}
&&f_6 = -  \frac{\epsilon\, c_1\,\sqrt{1+\beta r^2}}{r} + c_2\,,  \qquad 
f_3 = - \frac{c_1}{r} - c_1\,r\,\beta + \epsilon\, c_2 \sqrt{1 + \beta\,r^2}\,, \nonumber \\
&&f_5 = \frac{c_1\, \big(2 + r^2\, \beta +\epsilon\, \sqrt{1 + \beta\,r^2}\big) -c_2\,r\,\big(1 +\epsilon\, \sqrt{1 + \beta\,r^2}\big)}{r^3}\,,
\label{case4newdeteq}
\end{eqnarray}
where $c_1$ and $c_2$ are integration constants. Introducing (\ref{case4newdeteq}) 
back into the determining equations we get $c_1 = 0$ and then rest of the determining 
equations give us
\begin{eqnarray}
&&f_1 = -\frac{c_3}{\beta}\,\sqrt{1 + \beta\,r^2}\,, \qquad 
f_2 = \frac{2\hbar^2\,c_2\,\big(1 +\epsilon\, \sqrt{1 + \beta\, r^2}\big)}{r^2}\,, \nonumber \\
&&f_4 = \frac{c_3}{-\epsilon + \sqrt{1 + \beta\, r^2}}\,, \qquad 
V_0 = \hbar V_1\,.
\label{case5f1f2f4}
\end{eqnarray}

Finally, for this case, 
we have two arbitrary constants and hence two integrals of motion. One of them is a first-order 
operator and corresponds to the one already found in \cite{wy3} and the 
other is a new second-order operator. They are given as
\begin{eqnarray}
X_P^1 = -\frac{1}{\beta}\sqrt{1+\beta r^2}\,(\vec{\sigma}, \vec{p}) + 
\frac{(\vec{\sigma}, \vec{x})}{-\epsilon + \sqrt{1+\beta r^2}}\,\,  \Big((\vec{x}, \vec{p}) -i\hbar \Big)\,,
\label{case4intofmotfromoldart}
\end{eqnarray}
\begin{eqnarray}
X_P^2 &=& \frac{2}{r^2}\big(1+\epsilon\, \sqrt{1+\beta\,r^2}\big)\,(\vec{\sigma}, \vec{x})\, 
\Big(\hbar^2 + 3 i\hbar \,(\vec{x}, \vec{p}) - \big(\vec{x}, (\vec{x}, \vec{p})\,\vec{p}\big)\Big) \nonumber \\ 
&-& 2i\hbar \Big(2+\epsilon\, \sqrt{1+\beta\,r^2}\Big)\, (\vec{\sigma}, \vec{p}) 
+ 2\,(\vec{x}, \vec{p})\,(\vec{\sigma}, \vec{p}) + 2 \epsilon\, \sqrt{1+\beta\,r^2}\,(\vec{\sigma}, \vec{x})\,(\vec{p}, \vec{p})\,.
\label{case4intofmot}
\end{eqnarray}
Also note that we obtain an obvious integral by multiplying $X_P^1$ by $(\vec{\sigma}, \vec{L})$, however, not functionally 
independent of $X_P^1$ and $X_P^2$.

%%%%%%%%%%%%%%%%%%%%%%%%%%%%%%%%%%%%%%%%%%%%%%%%%%%%%%%%%%%%%%%%%%%%%%%%%%%%%%%%%%%%%%%%%%%%%%%%%%%%%%%%%%%%%%%%%%%%%%%%%%%%%%%%%%
{\bf Case V:} $V_1 = \frac{\hbar}{2r^2}\,\left(1 + \frac{2\epsilon}{\sqrt{1 + \beta\,r^2}} \right)$, $\beta \equiv 2 \lambda$

For this type of potential the set 
of determining equations given in 
(\ref{thirdten1})-(\ref{thirdten10}) together with 
(\ref{thirdf6}) give us 
\begin{eqnarray}
&&f_6 = - \frac{2\epsilon\,c_1\,\sqrt{1+\beta r^2}}{r} + c_2\,,  \qquad 
f_3 = - \frac{c_1}{r} - c_1\,r\,\beta + \frac{\epsilon}{2}\,c_2 \sqrt{1 + \beta\,r^2}\,, \nonumber \\
&&f_5 = \frac{2\,c_1\, \big(2 + r^2\, \beta + 2\epsilon\,\sqrt{1 + \beta\,r^2}\big) - c_2\,r\,\big(2 +\epsilon\, \sqrt{1 + \beta\,r^2}\big)}{2\,r^3}\,,
\label{case5newdeteq}
\end{eqnarray}
where $c_1$ and $c_2$ are integration constants. Introducing (\ref{case5newdeteq}) 
back into the determining equations we get $c_2 = 0$ and then rest of the determining 
equations imply
\begin{eqnarray}
&&f_1 = 0\,, \qquad f_4 = 0\,, \nonumber \\
&&f_2 = \frac{-\hbar^2c_1\,\big(1 + 2\,r^2\, \beta\big)\,\Big(1 +4\epsilon \sqrt{1+\beta\,r^2} + 4r^2\beta\, \big(1+\epsilon\, \sqrt{1+\beta r^2}\big) \Big) + 2r^4\,(1+\beta\,r^2)\,c_3}{2\,r^3\,(1 + \beta\,r^2)^2}\,, \nonumber \\
&&V_0 = \frac{\hbar^2}{8 r^2\left(1+\beta r^2\right)^2}\left(7+10 r^2 \beta +8\epsilon \left(1+\beta r^2\right)^{3/2}\right) - \frac{\alpha}{4\beta(1+\beta\,r^2)}\,.
\label{case5f2}
\end{eqnarray}

Finally, for this case, we have only one arbitrary constant ($c_3 = \alpha\, c_1$) and hence only one second-order operator given as
\begin{eqnarray}
X_P &=& \frac{2\,r^4\,\alpha\,(1 + \beta\,r^2) - \hbar^2(1 + 2\beta\,r^2)\,\Big(1 + 4\epsilon\sqrt{1+\beta\,r^2} + 4r^2\beta\,\big(1 +\epsilon\, \sqrt{1 + \beta\,r^2}\big)\Big)}{2\,r^3\,(1 + \beta\,r^2)^2} \,(\vec{\sigma}, \vec{x}) \nonumber \\
&-& \frac{2i\hbar}{r^3}\left(5 + 3 \beta\, r^2 - \frac{\epsilon}{\sqrt{1+\beta\,r^2}} + 6\,\epsilon\, \sqrt{1+\beta\,r^2}\right)\,(\vec{\sigma}, \vec{x})\,(\vec{x}, \vec{p}) \nonumber \\
&+& \frac{2i\hbar}{r}\left(1 + \beta\, r^2 - \frac{\epsilon}{\sqrt{1+\beta\,r^2}} + 4\,\epsilon\, \sqrt{1+\beta\,r^2}\right)\,(\vec{\sigma}, \vec{p}) \nonumber \\
&-& \frac{2}{r}\,(1 + \beta\,r^2)\,(\vec{\sigma}, \vec{x})\,(\vec{p}, \vec{p}) + \frac{2}{r^3}\, \Big(2 + \beta\, r^2 + 2\,\epsilon\, \sqrt{1+\beta\,r^2} \Big)\,(\vec{\sigma}, \vec{x})\,\big(\vec{x},(\vec{x}, \vec{p})\vec{p} \big) \nonumber \\
&-& \frac{4\epsilon}{r}\,\sqrt{1 + \beta\,r^2}\,(\vec{x}, \vec{p})\,(\vec{\sigma}, \vec{p})\,.
\label{case5intofmot}
\end{eqnarray}

%%%%%%%%%%%%%%%%%%%%%%%%%%%%%%%%%%%%%%%%%%%%%%%%%%%%%%%%%%%%%%%%%%%%%%%%%%%%%%%%%%%%%%%%%%%%%%%%%%%%%%%%%%%%%%%%%%%%%%%%%%%%%%%%%%%%%
%%%%%%%%%%%%%%%%%%%%%%%%%%%%%%%%%%%%%%%%%%%%%%%%%%%%%%%%%%%%%%%%%%%%%%%%%%%%%%%%%%%%%%%%%%%%%%%%%%%%%%%%%%%%%%%%%%%%%%%%%%%%%%%%%%%%
%%%%%%%%%%%%%%%%%%%%%%%%%%%%%%%%%%%%%%%%%%%%%%%%%%%%%%%%%%%%%%%%%%%%%%%%%%%%%%%%%%%%%%%%%%%%%%%%%%%%%%%%%%%%%%%%%%%%%%%%%%%%%%%%%%%%
\section{Vector integrals of motion} 
\label{comconditionsforva}
Let us take a linear combination of the independent vectors given in 
(\ref{vectors})
\begin{equation}
{\widetilde {\vec{X}}}_V=\sum_{j=1}^{15} f_j(r) \vec{V}_j\,, 
\label{linvectors}
\end{equation}
and fully symmetrize it as described in Section \ref{symmetsec}. 
The symmetric form of (\ref{linvectors}) is written as 
\begin{eqnarray}
\vec{X}_V &=&
 \vec{x}\, \Big(2f_1 - i\hbar\,(4f_7 + \frac{f_2'}{r} + r f_7') +
\big(2f_{11} - i\hbar\,(4f_{15} + \frac{f_{12}'}{r} + r f_{15}') \big)(\vec{\sigma},\vec{L}) \nonumber \\
& +&\big(2f_7 - i\hbar\,(\frac{2 f_5'}{r} + \frac{f_8'}{r} + 10 f_{13} + 2 r f_{13}') \big)(\vec{x},\vec{p}) \nonumber\\
& +& 2\,\big(f_{15}\,(\vec{x},\vec{p})(\vec{\sigma},\vec{L}) + f_5 \vec{p}^{\,\,2}+f_{13} (\vec{x},(\vec{x},\vec{p})\,\vec{p})\big) \Big)
\nonumber \\
& +&\Big(2\,\big(f_2 - i\hbar\,(f_5 + 2 f_8 + \frac{r f_8'}{2}) + f_{12}\,(\vec{\sigma},\vec{L}) + f_8\, (\vec{x},\vec{p})\big) \Big) \vec{p}
\nonumber \\
& +&(\vec{x} \wedge \vec{\sigma})\, \Big(2 f_3 + i\hbar\,(f_{11} - 4 f_9 - \frac{f_4'}{r} - r f_9') + 2 f_6\, \vec{p}^{\,\,2} 
\nonumber \\
& +& 2 f_{14}\,(\vec{x},(\vec{x},\vec{p})\,\vec{p})+ \big(2 f_9 - i\hbar\,(\frac{f_{10}' + 2 f_6'}{r} + 2 r f_{14}^{\prime}) \big)\, (\vec{x},\vec{p})\Big)
\nonumber \\
& -&\Big(2 f_4 - i\hbar\,\big(2 f_6 + 4 f_{10} + f_{12} + r f_{10}'\big) + 2 f_{10}\, (\vec{x},\vec{p}) \Big)\, (\vec{\sigma} \wedge \vec{p})\,.
\label{fullsymmetricvectors}
\end{eqnarray}

The requirement $[H, \vec{X}_V]=0$ gives us the determining 
equations for the vectors. The determining equations, obtained by 
equating the coefficients of the third-order terms to zero in the 
commutativity equation, are
\begin{eqnarray}
&&f_5 + f_8 = 0\,, \qquad f_8^{\prime} = 0\,, \qquad f_{13} = 0\,, \qquad f_{14} - f_{15} = 0\,, \nonumber \\
&&f_{10} + r^2 f_{14} + f_6 = 0\,, \qquad f_6^{\prime} - f_{12}^{\,\prime} =0\,, \qquad \hbar f_{15} + 2 f_6 V_1 = 0\,, 
\label{thirdordetvector}
\end{eqnarray}
together with the following two differential equations 
\begin{eqnarray}
&&\hbar \Big(f_6 V_1\Big)^{\,\prime} + 2 r f_6 V_1^2 = 0\,, \nonumber \\
&&\hbar f_6^{\,\prime} - 2r^2\, \Big(f_6 V_1\Big)^{\,\prime} - 4rV_1\,f_6 = 0\,.  
\label{thirdordetvector3}
\end{eqnarray}

It is obvious that $f_6 = 0$ is a solution of the 
system given in (\ref{thirdordetvector3}). For $f_6 \neq 0$,
(\ref{thirdordetvector3}) implies a compatibility 
condition for $V_1$ which reads as 
\begin{eqnarray}
6\hbar rV_1^2 - 4r^3V_1^3 + \hbar^2V_1^{\prime} = 0\,.
\label{oldeqforV1}
\end{eqnarray}
The above differential equation for $V_1$ had already been considered 
in \cite{wy3}. Depending on the solutions 
of this equation we have several cases. The general solution is given by 
\begin{equation}
V_1=\frac{\hbar}{2r^2} \left(1+\frac{\epsilon}{\sqrt{1+\beta r^2}}\right)\,,
\label{V1vec}
\end{equation}
where $\epsilon^2=1$. Note that $V_1=\frac{\hbar}{r^2}$ and $V_1=\frac{\hbar}{2r^2}$ 
are special solutions with $(\epsilon, \beta) = (1,0)$ and 
$(1, \infty)$, respectively. The case $V_1=\frac{\hbar}{r^2}$ is induced 
by a gauge transformation had already been considered 
thoroughly in \cite{wy3}. 

If we introduce the solutions (\ref{V1vec}) back into the 
system (\ref{thirdordetvector3}), we obtain differential 
equations for $f_6$ which can be solved. Hence, bearing in 
mind that $f_6 = 0$ is also a solution of the 
system (\ref{thirdordetvector3}), we have several cases to 
be considered separately. 

%%%%%%%%%%%%%%%%%%%%%%%%%%%%%%%%%%%%%%%%%%%%%%%%%%%%%%%%%%%%%%%%%%%%%%%%%%%%%%%%%%%%%%%%%%%%%%
\subsection{$V_1$ is a solution of (\ref{oldeqforV1})} 
\label{subseccase1}

%%%%%%%%%%%%%%%%%%%%%%%%%%%%%%%%%%%%%%%%%%%%%%%%%%%%%%%%%%%%%%%%%%%%%%%%%%%%%%%%%%%%%%%%%%%%%%%%%%
{{\bf Subcase I:} $V_1 = \frac{\hbar}{2 r^2}$ 

For this type of potential (\ref{thirdordetvector3}) implies 
$f_6 = c_1 r$. Thus, introducing this together with the 
relations given in (\ref{thirdordetvector}) into the determining equations 
obtained by equating the coefficients of the second-order 
terms to zero in the commutativity equation we get  
\begin{eqnarray}
&&f_2 = c_3 - r^2\,f_7\,, \qquad f_4 = c_4 - r^2\, f_9\,, \qquad 
f_7 = -\frac{c_1 }{2 r} + \frac{c_5}{2 r^2}\,, \qquad f_8 = -c_6\,, \nonumber \\
&&f_{9} = \frac{c_7}{r}\,, \qquad f_{11} = \frac{c_6}{2 r^2} + f_9\,, \qquad 
f_{12} = \frac{c_1 r}{2} + r^3 f_7^{\prime}\,, \qquad c_6 = 2 c_4\,,
\label{firtscasesub1detsec}
\end{eqnarray}
where $c_i$ ($i=3, \ldots, 7$) are integration constants. 

The determining equations obtained by equating the coefficients of the 
first- and zeroth-order terms to zero in the commutativity equation, give us 
\begin{eqnarray}
c_4 \Big(3\hbar^2 - 4 r^3 \big(2 V_0^{\prime} + r V_0^{\prime \prime}\big)\Big) = 0\,.
\label{firtscasesub1eqforV0}
\end{eqnarray}
Hence we have two possibilities: Either $c_4 = 0$, or 
$3\hbar^2 - 4 r^3 \big(2 V_0^{\prime} + r V_0^{\prime \prime}\big) = 0$, 
which has the following solution
\begin{eqnarray}
V_0 = \frac{3\hbar^2}{8 r^2} - \frac{\alpha}{r}\,,
\label{firtscasesub1solV0}
\end{eqnarray}
where $\alpha$ is an integration constant and we have set an irrelevant additive constant equal to zero.

%%%%%%%%%%%%%%%%%%%%%%%%%%%%%%%%%%%%%%%%%%%%%%%%%%%%%%%%%%%%%%%%%%%%%%%%%%%%%%%%%%%%%%%%%%%%%%
{\bf I$_1$: $V_0$ is given as in (\ref{firtscasesub1solV0})}  

Upon introducing $V_0$ back into the determining equations 
coming from first- and zeroth-order terms we have
\begin{eqnarray} 
f_1 = \frac{c_4 (2\hbar^2 - 4 \alpha r) + c_7 r}{2 r^2} \,, \qquad 
f_3 = - \frac{\hbar^2c_3 - 2\hbar^2 c_1 r - 4 c_3 \alpha r }{4 r^2}\,, \qquad 
c_5 = - c_3\,.
\label{firstcasesub11}
\end{eqnarray}

Then all the determining equations are satisfied. We have 
4 arbitrary constants ($c_1, c_3, c_4, c_7$). Thus, 
we have 4 integrals of motion which read
\begin{eqnarray}
&&\vec{X}_V^1 = - \big(\vec{\sigma}, \vec{L}\big)\,\vec{p} + \frac{3\hbar}{2}\vec{p} - \frac{\hbar\vec{x}}{2 r^2} (\vec{x}, \vec{p}) + \frac{i\hbar}{2}(\vec{\sigma} \wedge \vec{p}) + i\hbar^2 \frac{\vec{x}}{2 r^2} + \frac{4 \alpha r - \hbar^2}{4 r^2} (\vec{x} \wedge \vec{\sigma})\,,
\label{firstcasesub11finint1} \\
&&\vec{X}_V^2 = 2 \vec{x} \vec{p}^{\,\,2} - 2 (\vec{x}, \vec{p})\,\vec{p} - \frac{\hbar\vec{x}}{r^2} \big(\vec{\sigma}, \vec{L}\big) + 2 i\hbar \vec{p} - \hbar(\vec{\sigma} \wedge \vec{p}) + i\hbar^2 \frac{(\vec{x} \wedge \vec{\sigma})}{2 r^2} + \frac{\vec{x}}{r^2} (\hbar^2-2\alpha r)\,,
\label{firstcasesub11finint2} \\ 
&&\vec{X}_V^3 = \frac{(\vec{x}, \vec{\sigma})}{r} \vec{L} + \frac{\hbar}{2 r} \big(\vec{x} - i(\vec{x} \wedge \vec{\sigma})\big)\,,
\label{firstcasesub11finint3} \\
&&\vec{X}_V^4 = r \vec{L} (\vec{\sigma},\vec{p}) + \frac{\hbar r}{2}\vec{p} - \frac{i\hbar r}{2} (\vec{\sigma} \wedge \vec{p}) + \vec{X}_V^3\,\Big(i\hbar  - (\vec{x}, \vec{p})\Big)\,.
\label{firstcasesub11finint4}
\end{eqnarray}
However, the vector integrals of motion (\ref{firstcasesub11finint3}) and (\ref{firstcasesub11finint4}) 
can be written as anticommutators of the pseudoscalar integrals of motion given in (\ref{pscase1int1}) 
and (\ref{pscase1int2}), respectively, \emph{i.e.},
\begin{eqnarray}
\{\vec{J}, X_P^1\} = 2 \vec{X}_V^3\,, \quad {\rm and} \quad \{\vec{J}, X_P^2\} = - 2 \vec{X}_V^4\,.
\label{lasttrivialveciom}
\end{eqnarray}

%%%%%%%%%%%%%%%%%%%%%%%%%%%%%%%%%%%%%%%%%%%%%%%%%%%%%%%%%%%%%%%%%%%%%%%%%%%%%%%%%%%%%%%%
{\bf I$_2$: $V_0$ unspecified, $c_4 = 0$}  

From the determining equations obtained by equating the 
coefficients of the first- and zeroth-order terms to zero in the 
commutativity equation, we get 
\begin{eqnarray}
c_5 = -c_3\,, \qquad 
f_1 = \frac{c_7}{2 r} \,, \qquad c_3 \Big(3 \hbar^2 - 4 r^3 \big(2 V_0^{\prime} + r V_0^{\prime \prime}\big)\Big) = 0\,.
\label{firtscasesub12firstzerodeteq}
\end{eqnarray}

Thus, we must have $c_3=0$, because the other possibility gives us 
the previous potential for $V_0$. Upon introducing back all 
information into the determining equations we obtain
\begin{eqnarray}
f_3 = \frac{c_1}{2 r}\,,
\label{firtscasesub12lastdeteq}
\end{eqnarray}
and then all the determining equations are satisfied for 
arbitrary values of $V_0(r)$. We have two integrals of motion 
for this case which are given as in (\ref{firstcasesub11finint3}) 
and (\ref{firstcasesub11finint4}). It is worth noting that 
the commutativity condition is satisfied for arbitrary scalar potentials 
$V_0(r)$. 

%%%%%%%%%%%%%%%%%%%%%%%%%%%%%%%%%%%%%%%%%%%%%%%%%%%%%%%%%%%%%%%%%%%%%%%%%%%%%%%%%%%%%%%%%%%%%
{{\bf Subcase II:} $V_1 = \frac{\hbar}{2r^2} \left(1+\frac{\epsilon}{\sqrt{1+\beta r^2}}\right)$ 

For this type of potential (\ref{thirdordetvector3}) implies 
$f_6 = c_2 \sqrt{1+\beta r^2}$. Thus, introducing this together 
with the relations given in (\ref{thirdordetvector}) into the 
determining equations obtained by equating the coefficients 
of the second-order terms to zero in the commutativity 
equation we get 
\begin{eqnarray}
&&f_2 = c_3 - r^2\,f_7\,, \qquad f_4 = c_4 - r^2\, f_{11}\,, \qquad f_8 = 0\,, \qquad f_9 = f_{11}\,, \nonumber \\
&&f_{11} = \frac{c_4 + \epsilon\, c_4 \sqrt{1 + \beta r^2}}{r^2}\,, \qquad 
f_{12} = \frac{\epsilon\, r^2 (1 + \beta r^2) \left(c_2 \beta - 2\epsilon (1+\beta r^2) f_7\right)}{1+\epsilon\, \sqrt{1+\beta r^2} + \beta r^2 \left(2+\beta r^2 + 2\epsilon \sqrt{1+\beta r^2}\right)}\,,
\label{detmissing2011}
\end{eqnarray}
where $c_3$ and $c_4$ are integration constants. 

Upon introducing the relations given in (\ref{detmissing2011}) 
into the remaining determining equations, we have 
\begin{eqnarray}
f_1 = 0\,,
\end{eqnarray}
together with $c_2 = 0$ and $c_4 = 0$. However, $c_2 = 0$ reduces $f_6$ to zero 
which will be analyzed below. 

%%%%%%%%%%%%%%%%%%%%%%%%%%%%%%%%%%%%%%%%%%%%%%%%%%%%%%%%%%%%%%%%%%%%%%%%%%%%%%%%%%%%%%%%%%%%%%%%%%%%%%%%%%%%%%%
\subsection{$V_1$ unspecified, $f_6 = 0$} 
\label{subseccase2}
For this case let us continue to analyze the determining equations obtained 
by equating the coefficients of the second-order terms to zero in the 
commutativity equation. Introducing $f_6 = 0$ 
together with the relations given in (\ref{thirdordetvector}) into 
the determining equations we obtain 
\begin{eqnarray}
&&f_2 = c_3 - r^2\,f_7\,, \qquad f_4 = c_4 - r^2\, f_9\,, \qquad f_7 = f_{12}\,(V_1 + r V_1^{\prime})\,, \nonumber \\
&&f_{9} = \frac{2\hbar c_4 + \hbar f_8}{2r^2\,V_1 - \hbar}\,, \qquad f_{11} = f_9 + f_8\,(V_1 + r V_1^{\prime})\,,
\label{detsecvec}
\end{eqnarray}
together with the relations 
\begin{eqnarray}
&&2r(2c_4 + f_8)\,(2r^2V_1 - 3\hbar)\,V_1^2 + 2\Big(f_8\big(\hbar^2+6r^2V_1\,(r^2V_1 - \hbar)\big) - \hbar^2c_4\Big)\,V_1^{\prime} + rf_8(\hbar-2r^2V_1)^2\,V_1^{\prime \prime} = 0\,, \nonumber \\
&&2r(2c_4 + f_8)\,(2r^2V_1 - 3\hbar)\,V_1^2 - 2\Big(f_8\big(2\hbar^2+6r^2V_1\,(r^2V_1 - \hbar)\big) + \hbar^2c_4\Big)\,V_1^{\prime} - rf_8(\hbar-2r^2V_1)^2\,V_1^{\prime \prime} = 0\,, \nonumber \\
&&(2c_4 + f_8)\, (-6\hbar rV_1^2 + 4r^3V_1^3 - \hbar^2V_1^{\prime}) = 0\,, \qquad 
f_{12}\,(3 V_1^{\prime} + r V_1^{\prime \prime}) + f_{12}^{\prime}\,(V_1 + r V_1^{\prime})= 0\,, \nonumber \\
&&f_8\,(3 V_1^{\prime} + r V_1^{\prime \prime}) =0\,.
\label{detsecvectwo}
\end{eqnarray}
Here $c_3$ and $c_4$ are arbitrary constants of integration. From the 
system (\ref{detsecvectwo}) we have 4 subcases.

%%%%%%%%%%%%%%%%%%%%%%%%%%%%%%%%%%%%%%%%%%%%%%%%%%%%%%%%%%%%%%%%%%%%%%%%%%%%%%%%%%%%%%%
{\bf Subcase I: $f_8=-2c_4\,, \quad f_{12}=c_7\,, \quad 3V_1'+rV_1''=0\,,$}
which implies
\begin{eqnarray}
V_1 = \frac{\alpha}{2r^2} + \beta\,.
\label{newpotcase1}
\end{eqnarray}
This is a new spin-orbital potential. 

When we introduce all the information we have found up to 
now into the remaining determining equations, we get 
\begin{eqnarray}
&&c_4\,\Big(\alpha^2 - \big(4\,\beta\,r^2(\beta\,r^2 -\hbar)\big)\Big) + r^2\,
\Big(2\hbar^2f_1 + 4\,c_4 r \,V_0^{\prime}\Big) = 0\,, \\
&&c_7 = c_3\,, \qquad c_4\, \Big(4\,r\,V_0^{\prime \prime} + 8V_0^{\prime} -12\, \beta^2 r - 
\frac{\alpha(\alpha + 2 \hbar)}{r^3} \Big) =0\,. 
\label{detfirstzerocase1}
\end{eqnarray}

From equation (\ref{detfirstzerocase1}) it is immediately seen that 
\begin{eqnarray}
4\,r\,V_0^{\prime \prime} + 8V_0^{\prime} -12\, \beta^2 r - 
\frac{\alpha(\alpha + 2 \hbar)}{r^3} =0\,, 
\label{diffeqforV0case1vec}
\end{eqnarray}
because the other possibility $c_4 = 0$, either gives us 
the known potentials 
($V_1 = \frac{\hbar}{r^2}$ and $V_1 = \frac{\hbar}{2 r^2}$)
or reduces the number of the integral of motions. The 
solution of (\ref{diffeqforV0case1vec}) is given as 
\begin{eqnarray}
V_0 = \frac{\alpha(\alpha+2 \hbar) + 4\beta^2\,r^4 - 8r \gamma}{8\,r^2}\,,
\label{solforV0case1vec}
\end{eqnarray}
where $\gamma$ is an integration constant. Then, introducing back 
the solution given in (\ref{solforV0case1vec}) into the 
remaining determining equations, we obtain
\begin{eqnarray}
f_{1} = \frac{2\,c_4\big(\hbar\alpha - 2\,r\,(\gamma + \hbar\beta r)\big)}{2 r^2}\,, \qquad 
f_3 = \frac{c_3\,\Big(4r(\gamma + \beta^2 r^3) - \alpha^2\Big)}{4\,r^2\,(\alpha + 2\beta r^2)}\,, \qquad
\gamma = 0\,.
\label{detfirstzerocase12}
\end{eqnarray}

For these values of $f_j$, ($j = 1, \ldots, 15$) all the determining 
equations are satisfied for the potentials (\ref{newpotcase1}) and (\ref{solforV0case1vec}). We have two arbitrary constants, and hence we have two different integrals of motion. They are given as
\begin{eqnarray}
\label{finintofmotionveccase1a}
&&\vec{X}_V^1 = \vec{x}\,\Big(2\vec{p}^{\,2} - \left(\frac{2\beta r^2 -\alpha}{r^2}\right) \Big(\big(\vec{\sigma}, \vec{L}\big) + \hbar\Big) \Big) 
+ 2 \big(i\hbar - (\vec{x}, \vec{p})\big)\,\vec{p} - \hbar(\vec{\sigma} \wedge \vec{p}) \nonumber \\
&&\qquad \,\,+ i\hbar \left(\frac{2\beta r^2 -\alpha}{2 r^2}\right)\,(\vec{\sigma} \wedge \vec{x})\,,  \\
&&\vec{X}_V^2 = \Big(\frac{1}{2} (2 \hbar +\alpha -2 \beta r^2) + \big(\vec{\sigma}, \vec{L}\big) \Big)\,\vec{p} + \frac{\vec{x}}{2}\left(\frac{2\beta r^2 -\alpha}{r^2}\right) \Big((\vec{x}, \vec{p}) - i\hbar\Big) + \frac{i\hbar}{2} (\vec{\sigma} \wedge \vec{p}) \nonumber \\
&& \qquad - \frac{\hbar}{4}\left(\frac{2\beta r^2 -\alpha}{r^2}\right) (\vec{\sigma} \wedge \vec{x})\,. 
\label{finintofmotionveccase1} 
\end{eqnarray}

%%%%%%%%%%%%%%%%%%%%%%%%%%%%%%%%%%%%%%%%%%%%%%%%%%%%%%%%%%%%%%%%%%%%%%%%%%%%%%%%%%%%%%%%%%%%%%%%%
{\bf Subcase II: $f_8=0\,, \quad f_{12} = 0\,, \quad 6 \hbar rV_1^2-4r^3V_1^3+\hbar^2 V_1'=0\,.$} 

In this subcase depending on the solutions of (\ref{oldeqforV1}) we have 
two possibilities.

%%%%%%%%%%%%%%%%%%%%%%%%%%%%%%%%%%%%%%%%%%%%%%%%%%%%%%%%%%%%%%%%%%%%%%%%%%%%%%%%%%%%%
{\bf II$_1$:} $V_1 = \frac{\hbar}{2 r^2}$ 

Since for this type of potential $f_{9}$ in (\ref{detsecvec}) becomes undefined, we need to 
analyze this case from the beginning. Analysis of the determining equations obtained 
by equating the coefficients of the second-order terms to zero in the commutativity equation gives 
\begin{eqnarray}
f_2 = c_3\,, \qquad f_4 = -r^2 f_9\,, \qquad f_7 = 0\,, \qquad f_{11} = f_9\,, \qquad
f_{9} = \frac{c_4}{r}\,,
\label{case2subcase2caseifin}
\end{eqnarray}
where $c_3$ and $c_4$ are integration constants. For this case we already have 
$f_8 = 0$ and $f_{12} = 0$. From the other determining equations, we have 
\begin{eqnarray}
f_1 = \frac{c_4}{2 r}\,, \qquad 
f_3 = 0\,, \qquad c_3 = 0\,.
\label{case2subcase2caseifin2}
\end{eqnarray}

All the determining equations are satisfied for arbitrary potentials 
$V_0(r)$ and since we have only one arbitrary constant, namely $c_4$, there is 
only one integral of motion for this case. It is given by the equation (\ref{firstcasesub11finint3}) 
and is first-order in the momenta.

%%%%%%%%%%%%%%%%%%%%%%%%%%%%%%%%%%%%%%%%%%%%%%%%%%%%%%%%%%%%%%%%%%%%%%%%%%%%%%%%%%%%%%%%%
{\bf II$_2$:} $V_1 = \frac{\hbar}{2r^2} \left(1+\frac{\epsilon}{\sqrt{1+\beta r^2}}\right)$ 

Upon introducing all the information we have gathered so far into the determining equations 
obtained from equating the first- and zeroth-order terms to zero in the commutativity equation 
we find 
\begin{eqnarray}
f_1 = \frac{c_4}{2 r^2}\left(2 + \epsilon \frac{2 + \beta r^2 (4 + \beta r^2)}{(1 + \beta r^2)^{\frac{3}{2}}} \right)\,, \qquad 
f_3 = - \frac{c_3 \big(1 + \epsilon\, \sqrt{1 + \beta r^2}\big)}{2 r^2}\,.
\label{case2subcase2caseiifin}
\end{eqnarray}

If we introduce the above values of $f_1$ and $f_3$ back into the determining equations we see that we must have $c_3=0$ and $c_4=0$. Then all the determining equations are satisfied. Hence, we do not have any integral of motion for this case except the obvious one which is the anticommutator of the pseudoscalar integral of motion (\ref{case4intofmotfromoldart}) with $\vec{J}$, namely; 
\begin{eqnarray}
\vec{X}_V = \{X_P^1, \vec{J}\}\,.
\label{revisedadditionaliomvector}
\end{eqnarray}

%%%%%%%%%%%%%%%%%%%%%%%%%%%%%%%%%%%%%%%%%%%%%%%%%%%%%%%%%%%%%%%%%%%%%%%%%%%%%%%%%%%%%%%%%%%%%
{\bf Subcase III:} 

In this subcase we have $f_8 = 0$, $c_4 = 0$ and $f_{12} = 0$.
Upon introducing this information into the determining equations 
obtained from the first- and zeroth-order terms in the commutativity 
equation we have
\begin{eqnarray}
f_1 = 0\,, \qquad 
f_3 = \frac{c_3 V_1^{\prime}}{2 r V_1}\,, \qquad 
V_1 = \frac{\hbar + \epsilon \hbar\sqrt{1 + 4 C r^2}}{2 r^2}\,,
\label{fzcase2subcase3casei}
\end{eqnarray}
where $C$ is an integration constant. Now if we introduce these back into the determining equations we see that we either have $c_3 = 0$ or $C=0$. If $c_3=0$ we have no integral of motion 
since $c_4$ is already zero. If $C=0$ then we have $V_1 = \frac{\hbar}{2r^2}$ which has been investigated thoroughly. Hence, we conclude that no new information is obtained from this case. 

%%%%%%%%%%%%%%%%%%%%%%%%%%%%%%%%%%%%%%%%%%%%%%%%%%%%%%%%%%%%%%%%%%%%%%%%%%%%%%%%%%%%%%%%%%%%%%%
{\bf Subcase IV:} 
\begin{equation}
f_{12}\,(3 V_1^{\prime} + r V_1^{\prime \prime}) 
+ f_{12}^{\prime}\,(V_1 + r V_1^{\prime})= 0
\label{vecsb4}
\end{equation}

If we solve the above equation for $f_{12}$ and introduce back into the determining 
equations obtained from first- and zeroth-order terms in the commutativity equation 
we have 
\begin{eqnarray}
V_1 (V_1 + r V_1^{\prime})(3 V_1^{\prime} + r V_1^{\prime \prime})\,f_{12}^{\prime \prime} = 0\,.
\label{detforcase2subcase3caseii}
\end{eqnarray}

Thus we either have $V_1 + r V_1^{\prime} = 0$ or $f_{12}^{\prime \prime} = 0$. The other choice $3 V_1^{\prime} + r V_1^{\prime \prime} = 0$ has already been investigated. 
 
$V_1 + r V_1^{\prime} = 0$ implies $V_1 = \frac{C}{r}$, where $C$ is an integration constant. Then 
(\ref{vecsb4}) becomes 
\begin{eqnarray}
C\frac{f_{12}}{r^2} = 0\,.
\label{case2subcase3caseiifin}
\end{eqnarray}
Hence, we either have $C=0$ but then $V_1 = 0$ or $f_{12} = 0$ which has been investigated in the previous case. 

The condition $f_{12}^{\prime \prime} = 0$ implies 
\begin{eqnarray}
f_{12} = c_{10} r + c_{11}\,,
\label{case2subcase3caseiifin2}
\end{eqnarray}
where $c_{10}$ and $c_{11}$ are integration constants. Introducing this back into the determining equations we see that we have $c_{10} = 0$ and then (\ref{vecsb4}) becomes 
\begin{eqnarray}
c_{11} (3 V_1^{\prime} + r V_1^{\prime \prime}) = 0\,. 
\label{case2subcase3caseiifin3}
\end{eqnarray}
The two cases $c_{11}=0$ and $3 V_1^{\prime} + r V_1^{\prime \prime}=0$ have been previously investigated. 

Hence, we conclude that no new information is obtained from this case.

%%%%%%%%%%%%%%%%%%%%%%%%%%%%%%%%%%%%%%%%%%%%%%%%%%%%%%%%%%%%%%%%%%%%%%%%%%%%%%%%%%%%%%%%%%%%%%%%%%%%%%%%%%%%%%%%%%%%%%%%%%%%%%%%%%%%
%%%%%%%%%%%%%%%%%%%%%%%%%%%%%%%%%%%%%%%%%%%%%%%%%%%%%%%%%%%%%%%%%%%%%%%%%%%%%%%%%%%%%%%%%%%%%%%%%%%%%%%%%%%%%%%%%%%%%%%%%%%%%%%%%%%%
%%%%%%%%%%%%%%%%%%%%%%%%%%%%%%%%%%%%%%%%%%%%%%%%%%%%%%%%%%%%%%%%%%%%%%%%%%%%%%%%%%%%
\section{Axial vector integrals of motion}
\label{subsecaxialvectors}
Let us take a linear combination of the axial vectors given in
(\ref{axialvectors})
\begin{equation}
{\widetilde {\vec{X}}}_A=\sum_{j=1}^{15} f_j(r) \vec{A}_j\,,
\label{linaxialvectors}
\end{equation}
and fully symmetrize it as described in Section \ref{symmetsec}.
The symmetric form of (\ref{linaxialvectors}) is
\begin{eqnarray}
\vec{X}_A &=& \Big( (\vec{\sigma} , \vec{x}) \big( 2 f_{11}- i \hbar (2 f_{12} + f_6 + \frac{1}{r}f'_9 )\big) + 2 f_9 (\vec{\sigma},\vec{p})  \Big) \vec{p} \nonumber \\
& + & \Big(2 f_2 - i\hbar (5 f_5+r f_5')+2 f_5 (\vec{x},\vec{p}) +2 f_6 (\vec{L},\vec{\sigma}) \Big) \vec{L} + \vec{\sigma} \Big(2f_1-i\hbar \big( f_{11}+3f_4+f_8+rf'_4 \big) \nonumber \\
& + & \big(2f_4-i\hbar\big(f_{15}-2f_6+8f_7+\frac{2}{r}f'_3+2rf'_7\big) \big)(\vec{x},\vec{p})+2f_7 (\vec{x},(\vec{x},\vec{p}) \vec{p})+2f_3\vec{p}^{\,\,2} \Big) \nonumber \\
& + & \vec{x} \Big( (\vec{\sigma},\vec{x}) \big( 2f_{10}-i\hbar\big(5f_{13}+\frac{1}{r}(f_8' + f'_{11}) + rf'_{13}\big) + 2f_{12}\vec{p}^{\,\,2}+2f_{14}(\vec{x},(\vec{x},\vec{p}) \vec{p}) \big)   \nonumber \\
& + & \big(2f_8-i\hbar\big(2f_{12}+5f_{15}+f_6+\frac{1}{r}f'_9+rf'_{15}\big) \big) (\vec{\sigma},\vec{p}) \nonumber \\
& + & \big( 2f_{13}-i\hbar\big(12f_{14} + \frac{1}{r}(2 f'_{12} + f'_{15})+2rf'_{14}\big) \big) (\vec{\sigma},\vec{x}) (\vec{x},\vec{p})+ 2 f_{15} (\vec{x},\vec{p}) (\vec{\sigma},\vec{p}) \Big)\,. 
\label{fullsymmetricaxialvectors}
\end{eqnarray}

The determining equations obtained by
equating the coefficients of the third-order terms to zero in the
commutativity equation are
\begin{eqnarray}
&& f_5 = 0\,,  \\
\label{axial2}
&& 2r f_7+f_3'+r^2 f_7' = 0\,,  \\
\label{axial3}
&& 2r f_7+f_3'+r^2 f_6' = 0\,,  \\
\label{axial4}
&& \hbar f_{12} + 2 (f_9+f_3) V_1 = 0\,,  \\
&& 2 f_3 V_1+2 r^2 f_7 V_1-(f_{15}+f_{12}+r^2 f_{14}) (\hbar-2 r^2 V_1)-\hbar r (f_{15}'+f_6') = 0\,, \\
&& 2r f_{14} V_1 +\hbar f_{14}' = 0\,,  \\
&& \hbar f_{12}' + r (3\hbar f_{14}+2 (f_{12} + f_{15}+f_7) V_1) =0\,, \\
&& \hbar f_9'+r (\hbar f_{12}+\hbar f_{15}+2 f_3 V_1 - \hbar r f_6') = 0\,, \\
&& \hbar f_{12}+2 (f_9+f_3) V_1 + r (\hbar r f_{14}+2 r f_7 V_1 + \hbar f_6') = 0\,, \\
&& 2 r (f_{15}+r^2 f_{14}+2 f_7) V_1 - \hbar f_{15}' = 0\,. \label{axial11}
\end{eqnarray}

Equation (\ref{axial2}) can be integrated to give 
\begin{equation}
f_3=-r^2 f_7 - c_1\,,
\label{af3}
\end{equation}
where $c_1$ is a real constant. Introducing (\ref{af3}) into equation (\ref{axial3}) and integrating, we get 
\begin{equation}
f_7=f_6 + c_2\,,
\label{af7}
\end{equation}
where $c_2$ is an integration constant. We also solve (\ref{axial4}) 
for $f_{12}$ 
\begin{equation}
\hbar f_{12} = 2 (c_{1}-f_9+c_{2} r^2+f_{6} r^2) V_{1}\,.
\label{af8}
\end{equation}

With the obtained knowledge in equations (\ref{af3})-(\ref{af8}) we solve algebraically for derivatives of the functions.
Equations (\ref{axial2})-(\ref{axial11}) are then satisfied for arbitrary values of $V_1(r)$.

Satisfying the determining equations, obtained by
equating the coefficients of the third-order terms to zero in the
commutativity equation, for arbitrary values of $V_1(r)$ we continue 
with the determining equations, obtained by
equating the coefficients of the second-order terms to zero in the
commutativity equation. We note that four of the second-order 
determining equations involve only the functions $f_4, f_8, f_{11}$ and $f_{13}$: 
\begin{eqnarray}
f_{4} = 0\,,\quad (f_{11} - f_8) V_{1} = 0\,, \quad 
\hbar(f_{11}+f_{4}+f_8)-2r^2 V_{1} f_{11} = 0\,, \quad
\hbar f_{11}'+\hbar rf_{13} +2 r V_{1} (f_{11}+f_{4}) = 0\,.
\end{eqnarray}
They immediately imply that 
\begin{equation}
f_4 = 0\,, \quad f_8 = 0\,, \quad f_{11} = 0\,, \quad f_{13} = 0\,,
\end{equation}
which greatly simplifies the rest of the analysis. 

The complete set of determining equations is too 
long to present here. Instead, we present a nonlinear equation for $V_1$ which is obtained as a compatibility condition for $f_6$ in the analysis. Different 
solutions of this condition will form the different cases to be analyzed 
separately in which the full set of determining equations are fully 
investigated. The compatibility condition reads as 
\begin{eqnarray}
c_1 \frac{\Gamma_1 \Gamma_2}{\Gamma_3} = 0\,, \label{axialcombatforf6}
\end{eqnarray}
where
\begin{eqnarray}
&&\Gamma_1 = -20\hbar^2 r^4 V_{1}' V_1^3-60\hbar^2 r^3 V_1^4+16 r^7 V_1^6-5\hbar^4 r V_{1}'^{\,2} -3\hbar^4 V_{1} (V_{1}'-r V_{1}'')\,, \nonumber \\
&&\Gamma_2 = 3\hbar^5 V_{1}'+r \Big(-3\hbar^4 V_{1}'' (\hbar-r^2 V_{1}) +\hbar^2r \big(V_{1} V_{1}' (-\hbar^2 63+20 r^2 V_{1} (3\hbar-r^2 V_{1})) - 5\hbar^4 r V_1'^{\,2} \nonumber \\
&& \,\,\,\,\,\,\,\,\,\,+\,\,\,\, 4 r V_1^3 (-30\hbar^3+r^2 V_{1} (45\hbar^2-4 r^2 V_{1} (6\hbar-r^2 V_{1})))\big)\Big)\,, \nonumber \\
&&\Gamma_3 =  -3\hbar^4 r^2 V_{1}'' \big(2 V_{1} (3\hbar^2-6\hbar r^2 V_{1}+4 r^4 V_1^2) + \hbar^2r V_{1}' \big) + 6 r \Big(\hbar^2V_{1} V_{1}' \big(3\hbar^4+4 r^2 V_{1} (-9\hbar^3 \nonumber \\
&& \quad \,\,\,\, + \,\,\,\, 2 r^2 V_{1} (3\hbar^2+r^2 V_{1} (5\hbar-2 r^2 V_{1})))\big)+2 r^3 V_1^4 \big(-15\hbar^4+8 r^4 V_1^2 (9\hbar^2 - 2 r^2 V_{1} (4\hbar - r^2 V_{1}))\big) \nonumber \\
&& \,\,\,\,\,\,\,\,\,\,+\,\,\,\,  \hbar^4r V_{1}'^{\,2} \big(3\hbar^2 - 20 r^2 V_{1} (\hbar - r^2 V_{1})\big) \Big)\,.
\label{axialcombatintermsoffactors}
\end{eqnarray}

The compatibility condition (\ref{axialcombatforf6}) is 
satisfied if $c_1 = 0$ or $\Gamma_1 = 0$ or $\Gamma_2 = 0$. For $\Gamma_1 = 0$ or $\Gamma_2 = 0$ a standard symmetry analysis is performed in a similar fashion as for equation (\ref{compatibilityforv1}) and we find invariant solutions for $V_1$. 

The symmetry algebra for $\Gamma_1 = 0$ is spanned by two vector fields
\begin{equation}
\vec{v}_1 = r \partial_r - 2 V_1 \partial_{V_1} \,, \qquad \vec{v}_2 = r^3 \partial_r - 3 r^2 V_1 \partial_{V_1} \,,
\end{equation}
with $[\vec{v}_1 , \vec{v}_2] = 2 \vec{v}_2$. The invariant spin-orbit potentials are respectively
\begin{eqnarray}
V_1 = \frac{C_1}{r^2}\, \qquad {\rm and}  \qquad V_1 = \frac{C_2}{r^3}\,,
\label{ainvariantss}
\end{eqnarray}
where $C_1$ and $C_2$ are integration constants and only for the following
values 
\begin{eqnarray}
C_1 = \Big\{-\hbar,-\frac{\hbar}{2}, 0, \frac{\hbar}{2}, \hbar \Big\}\,,
\qquad {\rm and} \qquad C_2 = \big\{0\big\}\,,
\label{axialspecialsol1}
\end{eqnarray}
these two invariants are also solutions of $\Gamma_1 = 0$. Thus these values of 
the constants $C_1$ and $C_2$ give us the special solutions. The group transformations generated by $\vec{v}_2$ are
\begin{eqnarray}
\tilde{r} = \frac{r}{\sqrt{1 - 2 \lambda r^2}}\,, \qquad 
\widetilde{V}_1 = V_1 (1 - 2 \lambda r^2)^{\frac{3}{2}} \,, \qquad |\lambda| <\frac{1}{2r^2}\,.
\label{integratedgroup1}
\end{eqnarray}
Upon introducing the value of $V_1$ in the equation (\ref{integratedgroup1}), $\widetilde{V}_1$ becomes
\begin{eqnarray}
\widetilde{V}_1 = \dfrac{C_1}{\tilde{r}^2} \Big( \dfrac{1}{1 + 2 \lambda \tilde{r}^2} \Big)^\frac{1}{2}.
\label{integratedv11}
\end{eqnarray}
New special solutions of $\Gamma_1 = 0$ can be obtained from its
known solutions, say $V_1 = \frac{C_1}{r^2}$ with the constant $C_1$ taking
values from the set given in (\ref{axialspecialsol1}), since $\widetilde{V}_1(\tilde{r})$ is also
a solution if $V_1(r)$ is. Hence, we have the following solutions
\begin{eqnarray}
&&C_1=\, -\hbar \quad \Longrightarrow \quad \widetilde{V}_1 =
-\dfrac{\hbar}{\tilde{r}^2} \Big( \dfrac{1}{1 + 2 \lambda \tilde{r}^2} \Big)^\frac{1}{2} \,, \label{specialsol11}\\
&&C_1=-\frac{\hbar}{2} \quad \Longrightarrow \quad \widetilde{V}_1 = 
-\dfrac{\hbar}{2 \tilde{r}^2} \Big( \dfrac{1}{1 + 2 \lambda \tilde{r}^2} \Big)^\frac{1}{2}\,, \label{specialsol12} \\
&&C_1=\,\,\,\,\,0 \quad \Longrightarrow \quad \widetilde{V}_1 = 
0\,, \label{specialsol13} \\
&&C_1=\,\,\,\,\frac{\hbar}{2} \quad \Longrightarrow \quad \widetilde{V}_1 = 
\dfrac{\hbar}{2 \tilde{r}^2} \Big( \dfrac{1}{1 + 2 \lambda \tilde{r}^2} \Big)^\frac{1}{2} \,, \label{specialsol14} \\
&&C_1=\,\,\,\,\,\hbar \quad \Longrightarrow \quad \widetilde{V}_1 = 
\dfrac{\hbar}{\tilde{r}^2} \Big( \dfrac{1}{1 + 2 \lambda \tilde{r}^2} \Big)^\frac{1}{2} \,. \label{specialsol15} 
\end{eqnarray}
Together with the four solutions obtained from $V_1 = \frac{C_1}{r^2}$
with the constant $C_1$ taking values from the set given in (\ref{axialspecialsol1}),
we have eight different type of potentials $V_1$ for $\Gamma_1 = 0$.
%%%%%%%%%%%%%%%%%%%%%%%%%%%%%%%%%%%%%%%%%%%%%%%%%%%%%%%%%%%%%%%%%%%%%%%%%%%

The symmetry algebra for $\Gamma_2 = 0$ is spanned by two vector fields
\begin{equation}
\vec{v}_1 = r \partial_r - 2 V_1 \partial_{V_1} \,, \qquad \vec{v}_2 = r^3 \partial_r + (\hbar - 3 r^2 V_1) \partial_{V_1} \,,
\end{equation}
with $[\vec{v}_1 , \vec{v}_2] = 2 \vec{v}_2$. The invariants are
\begin{eqnarray}
V_1 = \frac{C_1}{r^2}\, \qquad {\rm and}  \qquad V_1 = \frac{C_2}{r^3}+\frac{\hbar}{r^2}\,,
\label{ainvariants2}
\end{eqnarray}
where $C_1$ and $C_2$ are integration constants and only for the following
values 
\begin{eqnarray}
C_1 = \Big\{0, \frac{\hbar}{2}, \hbar,\frac{3\hbar}{2},2\hbar \Big\}\,,
\qquad {\rm and} \qquad C_2 = \big\{0\big\}\,,
\label{axialspecialsol2}
\end{eqnarray}
these two invariants are also solutions of $\Gamma_2 = 0$.
Thus these values of the constants $C_1$ and $C_2$ give us the special
solutions. The group transformations induced by $\vec{v}_2$ are
\begin{eqnarray}
\tilde{r} = \frac{r}{\sqrt{1 - 2 \lambda r^2}}\,, \qquad 
\widetilde{V}_1 =\frac{1 - 2 \lambda r^2}{r^2} \Big( \hbar + (\hbar - r^2 V_1)\sqrt{1 - 2 \lambda r^2} \Big)\,, \qquad |\lambda| <\frac{1}{2r^2}\,.
\label{aintegratedgroup2}
\end{eqnarray}
Upon introducing the value of $V_1$ in the equation (\ref{aintegratedgroup2}), $\widetilde{V}_1$ becomes
\begin{eqnarray}
\widetilde{V}_1 = \dfrac{1}{\tilde{r}^2} \left(\hbar+(\hbar-C_1) \sqrt{\frac{1}{1 + 2 \lambda \tilde{r}^2}}  \right)\,.
\label{integratedv12}
\end{eqnarray}
New special solutions of $\Gamma_2 = 0$ can be obtained from its
known solutions, say $V_1 = \frac{C_1}{r^2}$ with the constant $C_1$ taking
values from the set given in (\ref{axialspecialsol2}), since $\widetilde{V}_1(\tilde{r})$ is also
a solution if $V_1(r)$ is. Hence, we have the following solutions
\begin{eqnarray}
&&C_1=\,\,\,\,\,0 \,\quad \Longrightarrow \quad \widetilde{V}_1 = 
\frac{\hbar}{\tilde{r}^2} \Big( 1 + \frac{1}{\sqrt{1 + 2 \lambda \tilde{r}^2}} \Big)\,, \label{specialsol21} \\
&&C_1=\,\,\,\,\,\frac{\hbar}{2} \quad \Longrightarrow \quad \widetilde{V}_1 = 
\frac{\hbar}{2 \tilde{r}^2} \Big( 2 + \frac{1}{\sqrt{1 + 2 \lambda \tilde{r}^2}} \Big)\,, \label{specialsol22} \\
&&C_1=\,\,\,\,\,\hbar \,\quad \Longrightarrow \quad \widetilde{V}_1 = 
\frac{\hbar}{\tilde{r}^2}\,, \label{specialsol23} \\
&&C_1=\,\,\,\,\,\frac{3\hbar}{2} \quad \Longrightarrow \quad \widetilde{V}_1 = 
\frac{\hbar}{2\tilde{r}^2} \Big( 2 - \frac{1}{\sqrt{1 + 2 \lambda \tilde{r}^2}} \Big)\,, \label{specialsol24} \\
&&C_1=\,\,\,\,\,2\hbar \,\quad \Longrightarrow \quad \widetilde{V}_1 = 
\frac{\hbar}{\tilde{r}^2} \Big( 1 - \frac{1}{\sqrt{1 + 2 \lambda \tilde{r}^2}} \Big)\,. \label{specialsol25}
\end{eqnarray}
Together with the two new solutions of the form $V_1 = \frac{C_1}{r^2}$,
we have six different type of potentials $V_1$ for this case.
%%%%%%%%%%%%%%%%%%%%%%%%%%%%%%%%%%%%%%%%%%%%%%%%%%%%%%%%%%%%%%%%%%%%%%%%%%%%%%%%%%%%%%%%%%%

Finally, if we consider the simultaneous 
solutions of $\Gamma_1 = 0$ and $\Gamma_2 = 0$ by substituting 
$V_1''$ from $\Gamma_1 = 0$ in $\Gamma_2 = 0$, we obtain 
\begin{equation}
6\hbar rV_1^2-4r^3V_1^3+\hbar^2 V_1' = 0.
\end{equation}
Solving this equation provides us with two new potentials
\begin{eqnarray}
V_1 = \frac{\hbar}{2r^2} \Big( 1 + \frac{\epsilon}{\sqrt{1+\beta r^2}} \Big)\,, \qquad \epsilon^2 = 1\,,
\end{eqnarray}
where $\beta$ is a constant of integration.
%%%%%%%%%%%%%%%%%%%%%%%%%%%%%%%%%%%%%%%%%%%%%%%%%%%%%%%%%%%%%%%%%%%%%%%%%%%%%%%

Although we have $16$ different type of potentials which 
satisfy the compatibility condition (\ref{axialcombatforf6}), 
we note that for the following potentials 
\begin{eqnarray}
V_1 = \frac{\epsilon\hbar}{2r^2}\,, \qquad 
V_1 = \frac{3\hbar}{2r^2}\,, \qquad 
V_1 = \frac{\epsilon\hbar}{2r^2 \sqrt{1 + 2 \lambda r^2}}\,, \nonumber \\
V_1 = \frac{\hbar}{2r^2} \left( 2 + \frac{\epsilon}{\sqrt{1 + 2\lambda r^2}} \right)\,, \qquad 
V_1 = \frac{\hbar}{2r^2} \left( 1 + \frac{\epsilon}{\sqrt{1 + \beta r^2}} \right)\,, 
\label{axialdisregpot}
\end{eqnarray}
$\Gamma_3$ becomes identically zero. Hence we do not consider these potentials 
in the rest of the analysis. Indeed if we consider them and start the investigation 
of the determining equations from the very beginning the analysis gives 
$c_1 = 0$. This possibility will be analyzed separately. Thus, excluding the 
potentials given in (\ref{axialdisregpot}) as well as the potential 
$V_1 = \frac{\hbar}{r^2}$, which is a gauge induced one and was completely 
investigated in \cite{wy3}, we now investigate the following five cases separately. 

%%%%%%%%%%%%%%%%%%%%%%%%%%%%%%%%%%%%%%%%%%%%%%%%%%%%%%%%%%%%%%%%%%%%%%%%%%%%%%%%%%%%%%%%%%%%%%%%%%%%%%%%%%%%%%%%%%%%%%%%%%%%%%%%%%
{\bf Case I:} $c_1 = 0$

For $c_1 = 0$ the third-order determining equations 
(\ref{axial2})-(\ref{axial11}) are satisfied by 
\begin{eqnarray}
f_7 = 0\,, \quad  
f_9 = 0\,, \quad
f_{12} = 0\,, \quad
f_{14} = 0 \,, \quad
f_{15} = 0\,, 
\end{eqnarray}
together with equations (\ref{af3}) and (\ref{af7}). Then 
the determining equations, obtained from lower order terms, 
provide us with
\begin{equation}
f_{1} = c_3\,, \qquad f_{2} = c_4\,, \qquad
f_{10} = 0\,, \qquad
c_4 = 2 c_3 - \frac{c_2}{2}\,, 
\end{equation}
where $c_3$ and $c_4$ are integration constants. For these values of 
$f_j$, all the determining equations, obtained from 
the requirement that the commutator $[H, \vec{X}_A]=0$, are satisfied for 
any $V_0 = V_0(r)$ and $V_1 = V_1(r)$. Since we have two arbitrary constants 
and none of them are appeared in the Hamiltonian we have the following 
integrals of motion  
\begin{eqnarray}
\vec{X}_A^1 &=& i\hbar\Big((\vec{\sigma},\vec{x})\,\vec{p} + \vec{x}\, (\vec{\sigma},\vec{p})\Big) -\hbar\vec{L} - 2 (\vec{\sigma},\vec{L}) \vec{L} - 
2 i\hbar \vec{\sigma}\, (\vec{x},\vec{p}) = \{(\vec{\sigma},\vec{L}),\vec{J}\} \,,  \label{axialveccase11} \\
\vec{X}_A^2 &=&  4 \vec{L} + 2\hbar \vec{\sigma} = 4 \vec{J} \label{axialveccase12} \,.
\label{axtrivec2}
\end{eqnarray}

They are however both trivial since $\vec{J}$ and $(\vec{\sigma},\vec{L})$ are in the set (\ref{trivialscalars}). From here on, we shall list only nontrivial integrals.

%%%%%%%%%%%%%%%%%%%%%%%%%%%%%%%%%%%%%%%%%%%%%%%%%%%%%%%%%%%%%%%%%%%%%%%%%%%%%%%%%%%%%%%%%%%%%%%%%%%%%%%%%%%%%%%%%%%%%%%%%%%%%%%%%%
{\bf Case II:} $V_1 = -\frac{\hbar}{r^2}$

For this type of potential, the set of determining equations 
given in (\ref{axial2})-(\ref{axial11}), is satisfied by 
\begin{eqnarray}
&&f_6 = -c_2\,, \qquad f_9 = 2 c_1 \,, \qquad f_{14} = 0\,, 
\qquad f_{15} = - \frac{4 c_1}{r^2} \,, 
\label{axialcase2third} 
\end{eqnarray} 
together with the equations given in (\ref{af3})-(\ref{af8}). 
Here, $c_1$ and $c_2$ are integration constants. If we introduce 
these integrals of motion into the set of all determining equations, we get 
\begin{eqnarray}
f_2 = c_3\,, \qquad f_{10} = 0\,, \qquad f_1 = \frac{c_2}{4} + \frac{c_3}{2} + c_4 r^2\,, 
\label{axialcase2lowodr}
\end{eqnarray}
where $c_3$ and $c_4$ are integration constants. Upon introducing 
all the information obtained so far back into the determining 
equations we find  
\begin{eqnarray}
V_0 = \alpha r^2\,,
\label{axialcase2V0sol}
\end{eqnarray}
with $\alpha = -\frac{c_4}{2c_1}$. Hence, 
the integral of motion $\vec{X}_A$ depends on three constants $c_1$, 
$c_2$ and $c_3$. The nontrivial integral of motion 
for this case can be written 
\begin{eqnarray}
\vec{X}_A & = &  -\big(2\alpha r^2 + \vec{p}^{\,\,2}\big)\vec{\sigma} + 2 (\vec{\sigma},\vec{p}) \vec{p} \nonumber \\
& + & \frac{2}{r^2}\Bigg(\vec{x} \Big((\vec{\sigma},\vec{x})\,\vec{p}^{\,\,2} + 2 i\hbar (\vec{\sigma},\vec{p}) - 2 (\vec{x},\vec{p})(\vec{\sigma},\vec{p})\Big) + i\hbar \big(\vec{\sigma} (\vec{x},\vec{p}) - (\vec{\sigma},\vec{x}) \vec{p}\,\big)\Bigg)\,.
\label{axialveccase23}
\end{eqnarray}

%%%%%%%%%%%%%%%%%%%%%%%%%%%%%%%%%%%%%%%%%%%%%%%%%%%%%%%%%%%%%%%%%%%%%%%%%%%%%%%%%%%%%%%%%%%%%%%%%%%%%%%%%%%%%%%%%%%%%%%%%%%%%%%%%%
{\bf Case III:} $V_1 = \frac{2\hbar}{r^2}$

For this type of potential, the set of determining equations 
given in (\ref{axial2})-(\ref{axial11}), is satisfied by 
\begin{eqnarray}
&&f_6 = -\frac{4 c_1}{r^2} - c_2\,, \qquad f_9 = - 2 c_1\,, \qquad f_{14} = \frac{8 c_1}{r^4} \,, \qquad f_{15} = 0\,, 
\label{axialcase3third} 
\end{eqnarray} 
together with the equations given in (\ref{af3})-(\ref{af8}). 
Here, $c_1$ and $c_2$ are integration constants. If we introduce 
these integrals of motion into the set of determining equations, 
obtained from lower order terms, we get 
\begin{eqnarray}
f_2 = -\frac{4 c_1}{r^2} + c_3\,, \qquad f_{10} = -\frac{6 c_1}{r^4} + c_4\,, \qquad f_1 = \frac{c_2}{4} + \frac{c_3}{2}  - \frac{c_4}{2} r^2\,, 
\label{axialcase3lowodr}
\end{eqnarray}
where $c_3$ and $c_4$ are integration constants. Upon introducing 
all the information obtained so far back into the determining 
equations we find  
\begin{eqnarray}
V_0 = \frac{3\hbar^2}{r^2} + \alpha r^2 \,,
\label{axialcase3V0sol}
\end{eqnarray}
with $\alpha = \frac{c_4}{4c_1}$. Hence, again 
the integral of motion $\vec{X}_A$ depends on three constants $c_1$, 
$c_2$ and $c_3$. Two of the integrals of motion are trivial and the third one is 
\begin{eqnarray}
\vec{X}_A &\!\!\!\!=\!\!\!\!& \big(3 \vec{p}^{\,\,2} - 2\alpha r^2+\frac{4}{r^2}\left(i\hbar (\vec{x},\vec{p})-(\vec{x},(\vec{x},\vec{p}) \vec{p})\right)\big)\vec{\sigma} -\frac{2}{r^2}\big(\hbar+(\vec{L},\vec{\sigma})\big)\vec{L} -2\left((\vec{\sigma},\vec{p})-\frac{3i \hbar(\vec{\sigma},\vec{x})}{r^2}\right)\vec{p} \nonumber \\ 
&\!\!\!\!+\!\!\!\!& \frac{2 \vec{x}}{r^4}\bigg(3i\hbar r^2(\vec{\sigma},\vec{p})-(\vec{\sigma},\vec{x})\big(3h^2-2r^4\alpha+2r^2\vec{p}^{\,2}+12i\hbar (\vec{x},\vec{p})\big)-4(\vec{x},(\vec{x},\vec{p}) \vec{p})\bigg) \,.
\label{axialveccase30}
\end{eqnarray}

%%%%%%%%%%%%%%%%%%%%%%%%%%%%%%%%%%%%%%%%%%%%%%%%%%%%%%%%%%%%%%%%%%%%%%%%%%%%%%%%%%%%%%%%%%%%%%%%%%%%%%%%%%%%%%%%%%%%%%%%%%%%%%%%%%
{\bf Case IV:} $V_1 = \frac{\epsilon\hbar}{r^2 \sqrt{1 + \beta r^2}}$, $\beta \equiv 2 \lambda$, $\epsilon^2=1$

For this type of potential, the set of determining equations
given in (\ref{axial2})-(\ref{axial11}), is satisfied by
\begin{eqnarray}
f_6 = -c_2-\frac{2c_1}{r^2}
\Big( 1+\frac{\beta}{2} r^2+\epsilon\, \sqrt{1+\beta r^2} \Big), \quad f_9 = - 2\epsilon\, c_1 \sqrt{1+\beta r^2}, \quad f_{14} = 0\,, \quad f_{15} = \frac{4 \epsilon\, c_1 \sqrt{1+\beta r^2}}{r^2} \,, 
\label{axialcase4third}
\end{eqnarray}
together with the equations given in (\ref{af3})-(\ref{af8}).
Here, $c_1$ and $c_2$ are integration constants. If we introduce
these integrals of motion into the set of determining equations,
obtained from lower order terms, we get
\begin{eqnarray}
f_1 =\frac{c_2}{4}+\frac{c_3}{2}+\frac{c_1}{4 r^2} \left(1+\frac{3}{\left(1+\beta r^2\right)^2}-\frac{8 \left(1+\alpha  r^2\right)}{1+\beta r^2}- \frac{4\epsilon}{\sqrt{1+\beta r^2}}\right) \,, \nonumber \\
f_2 = c_3 -\frac{2 c_1}{r^2} \left(1+2 \alpha r^2 + \frac{\epsilon}{\sqrt{1+\beta r^2}}\right)\,, \qquad f_{10} = 0 \,, 
\label{axialcase4lowodr}
\end{eqnarray}
where $c_3$ and $\alpha$ are integration constants. Upon introducing
all the information obtained so far back into the determining
equations we find
\begin{eqnarray}
V_0 = \frac{\hbar^2}{8 r^2 \left(1+\beta r^2\right)^2} \left( 4+6 \beta r^2 - r^4 \beta ^2 + 4 \epsilon\, \left(1+\beta r^2 \right)^{\frac{3}{2}}\right)+\frac{\alpha }{1+\beta r^2}\,,
\label{axialcase4V0sol}
\end{eqnarray}
Hence, again
the integral of motion $\vec{X}_A$ depends on three constants $c_1$,
$c_2$ and $c_3$. The only nontrivial integral of motion is
\begin{eqnarray}
\vec{X}_A &=& \left( 2 i\hbar \frac{(\vec{\sigma}, \vec{x})}{r^2} Q_+  - 4 \epsilon \sqrt{1+\beta r^2} \, (\vec{\sigma},\vec{p})  \right) \vec{p} - \frac{4}{r^2}\left(\hbar + 2 \alpha r^2 + \frac{\epsilon\hbar}{\sqrt{1+\beta r^2}} + q(\vec{\sigma},\vec{L})\right)\,  \vec{L} \nonumber \\
& + & \frac{\vec{\sigma}}{2 r^2} \Bigg(Y + 8 i\hbar \left(1+\frac{\beta}{2} r^2\right)(\vec{x},\vec{p}) - 8 q (\vec{x},(\vec{x},\vec{p}) \vec{p}) + 4 r^2 \left(2q-1\right)\vec{p}^{\,\,2} \Bigg) 
\nonumber \\ 
& + & \frac{2 \vec{x}}{r^2} \Bigg(\bigg( i\hbar Q_- + 4\epsilon \sqrt{1+\beta r^2} (\vec{x},\vec{p})\bigg) (\vec{\sigma},\vec{p}) - 2\epsilon (\vec{\sigma},\vec{x}) \sqrt{1+\beta r^2} \vec{p}^{\,\,2} \Bigg)\,,
\label{newaxialcase4intofmot3}
\end{eqnarray}
where $Q_{\pm}$, $q$ and $Y$ are given by the following relations
\begin{eqnarray}
Q_{\pm} = 1 + \frac{\beta}{2} r^2 \pm \epsilon\frac{3 + 4 \beta r^2}{\sqrt{1+\beta r^2}}\,, \qquad 
q = 1+\frac{\beta}{2} r^2+\epsilon\, \sqrt{1+\beta r^2}\,, \label{simplifyiomrevised1} \\
Y = \frac{-4 \hbar^2 - 8 r^2 \alpha - 6\hbar^2 r^2 \beta - 8r^4\alpha \beta + \hbar^2 r^4 \beta^2}{(1 + \beta r^2)^2} - \frac{4\epsilon \hbar^2}{\sqrt{1+\beta r^2}}\,.
\label{simplifyiomrevised2}
\end{eqnarray}

%%%%%%%%%%%%%%%%%%%%%%%%%%%%%%%%%%%%%%%%%%%%%%%%%%%%%%%%%%%%%%%%%%%%%%%%%%%%%%%%%%%%%%%%%%%%%%%%%%%%%%%%%%%%%%%%%%%%%%%%%%%%%%%%%%
{\bf Case V:} $V_1 = \frac{\hbar}{r^2} \left( 1 + \frac{\epsilon}{\sqrt{1 + \beta r^2}} \right)$, $\beta \equiv 2 \lambda$, $\epsilon^2=1$

For this type of potential, the set of determining equations
given in (\ref{axial2})-(\ref{axial11}), is satisfied by
\begin{eqnarray}
&&f_6 = -c_2-\frac{2c_1}{r^4}
\Big( 1+\frac{\beta}{2} r^2+\epsilon\, \sqrt{1+\beta r^2} \Big) \,, \qquad f_9 = - 2 \epsilon\, c_1 \sqrt{1+\beta r^2} \,, \nonumber \\
&&f_{14} = \frac{4 c_1}{r^4} \left(1+\frac{\beta}{2} r^2+\epsilon\, \sqrt{1+\beta r^2}\right) \,, \qquad f_{15} = 0 \,, 
\label{axialcase5third}
\end{eqnarray}
together with the equations given in (\ref{af3})-(\ref{af8}).
Here, $c_1$ and $c_2$ are integration constants. If we introduce
these integrals of motion into the set of determining equations,
obtained from lower order terms, we get
\begin{eqnarray}
f_1 =\frac{c_2}{4}+\frac{c_3}{2} - c_1\left(\frac{r^2 \left(4 \alpha -6 \beta ^2\right)+r^4 \left(3 \beta ^3+4 \beta \alpha \right)-6 \beta}{4 \left(1+\beta r^2\right)^2} -  \frac{2 \epsilon\, \beta }{\sqrt{1+\beta r^2}}\right), \nonumber \\
f_2 = c_3 -\frac{2 c_1}{r^2} \left(1+ \frac{\epsilon}{\sqrt{1+\beta r^2}}\right), \quad 
f_{10} = -\frac{c_1 }{r^4}\left(9+\frac{3}{2 \left(1+\beta r^2\right)^2}-\frac{15+4 r^4 \alpha}{2\left(1+\beta r^2\right)} + \frac{3\epsilon \left(1+2 \beta r^2\right)}{\sqrt{1+\beta r^2}}\right),
\label{axialcase5lowodr}
\end{eqnarray}
where $c_3$ and $\alpha$ are integration constants. Upon introducing
all the information obtained so far back into the determining
equations we find
\begin{eqnarray}
V_0 = \frac{3\hbar^2 \Big(4+5 \beta r^2+4\epsilon \left(1 + \beta r^2\right)^{\frac{3}{2}}\Big)}{8 r^2 \left(1+\beta r^2\right)^2}-\frac{\alpha }{2 \beta \left(1+\beta r^2\right)}\,.
\label{axialcase5V0sol}
\end{eqnarray}
Hence, again
the integral of motion $\vec{X}_A$ depends on three constants $c_1$,
$c_2$ and $c_3$. The only nontrivial integral of motion is
\begin{eqnarray}
\vec{X}_A &=& \Bigg( \frac{2 i\hbar^2 (\vec{\sigma} , \vec{x})}{r^2} \widetilde{Q} - 4 \epsilon\, \sqrt{1+\beta r^2}\, (\vec{\sigma},\vec{p})  \Bigg) \vec{p} - \frac{4}{r^2}\Bigg(\hbar + \frac{\epsilon\hbar}{\sqrt{1 + \beta r^2}} + q (\vec{\sigma}, \vec{L}) \Bigg) \vec{L} \nonumber \\
& + & \frac{2 \vec{\sigma}}{r^2} \Bigg(2 i\hbar q (\vec{x},\vec{p}) - \widetilde{Y}  - 2 q (\vec{x},(\vec{x},\vec{p}) \vec{p}) + r^2 \left(2q-1\right)\vec{p}^{\,\,2} \Bigg) 
+ \frac{2 \vec{x}}{r^4} \Bigg((\vec{\sigma},\vec{x}) \Big(4 q (\vec{x},(\vec{x},\vec{p}) \vec{p}) \nonumber \\
& - & 2 r^2 \left(2q-1-\epsilon\, \sqrt{1+\beta r^2}\right) \vec{p}^{\,\,2} -
Z - 4 i\hbar W (\vec{x},\vec{p})\Big) + i\hbar \widetilde{Q}(\vec{\sigma},\vec{p})  \Bigg)\,,
\label{newaxialcase5intofmot3}
\end{eqnarray}
where $q$ is given in (\ref{simplifyiomrevised1}) and $\widetilde{Q}$, $\widetilde{Y}$, 
$Z$ and $W$ are given by the following relations
\begin{eqnarray}
\widetilde{Q} = 3 + \frac{5 \beta}{2} r^2 + \epsilon\frac{3 + 4 \beta r^2}{\sqrt{1+\beta r^2}}\,, \qquad 
W = 3+ 2 \beta r^2 + \epsilon\, \frac{6 + 7 \beta r^2}{2 \sqrt{1+\beta r^2}}\,, \\
\widetilde{Y} = \frac{\hbar^2 r^4 (4 \alpha - 6 \beta^2) + \hbar^2 r^6(3 \beta^3 + 4\beta \alpha) -6 \hbar^2 \beta r^2}{4 (1 + \beta r^2)^2} + \frac{2\epsilon \hbar^2 \beta r^2}{\sqrt{1+\beta r^2}}\,, \\
Z = \frac{-4 r^4 \alpha (1 + \beta r^2) + 3 \hbar^2 \big(2 + \beta r^2 (7 + 6 \beta r^2)\big)}{2 (1 + \beta r^2)^2} + \epsilon \frac{3 \hbar^2 (1 + 2 \beta r^2)}{\sqrt{1 + \beta r^2}}\,.
\end{eqnarray}

%%%%%%%%%%%%%%%%%%%%%%%%%%%%%%%%%%%%%%%%%%%%%%%%%%%%%%%%%%%%%%%%%%%%%%%%%%%%%%%%%%%%%%%%%%%%%%%
%%%%%%%%%%%%%%%%%%%%%%%%%%%%%%%%%%%%%%%%%%%%%%%%%%%%%%%%%%%%%%%%%%%%%%%%%%%%%%%%%%%%%%%%%%%%%%%
%%%%%%%%%%%%%%%%%%%%%%%%%%%%%%%%%%%%%%%%%%%%%%%%%%%%%%%%%%%%%%%%%%%%%%%%%%%%%%%%%%%%%%%%%%%%%%%
\section{Conclusions}
The main results of this article are summed up in Table 1. We list all potentials $V_0(r)$ and $V_1(r)$ that have at least one integral of motion in addition to the ``trivial'' ones, \emph{i.e.} those that exist for all $V_0(r)$ and $V_1(r)$ (see equation (\ref{trivialscalars})). We have included the obvious integrals obtained by multiplying a lower-order integral by the trivial ones. 

The results are ordered according to the value of $V_1(r)$ (in column 2). In column 3, $V_0(r)$ denotes an arbitrary function. Throughout, $\alpha$ and $\beta$ are real arbitrary constants, and we have $\epsilon=\pm 1$. The nontrivial pseudoscalar, vector and axial vector integrals are listed in columns 4,5 and 6. We have included integrals of order 0 and 1 (in the momenta) already found in \cite{wy3}. Indeed, the potentials in row 1 are first order superintegrable.

We mention that the nonlinear ordinary differential equations (\ref{compatibilityforv1}) and $\Gamma_1=0$, $\Gamma_2=0$ with $\Gamma_1$ and $\Gamma_2$ defined in (\ref{axialcombatintermsoffactors}), are compatibility conditions for the existence of pseudoscalar and axial vector integrals, respectively. We have obtained the general solutions of these equations in implicit form. Since a potential must be explicit in order to be useful, we only used the particular explicit solutions obtained by requiring that they be invariant under a subgroup of the symmetry groups of these equations. Whenever possible, we enlarged the class of solutions by acting on the subgroup invariant solution with the entire symmetry group of the auxiliary equations.

We consider the most interesting cases to be those with vector or axial vector integrals, specially those that can be viewed as a Coulomb atom or harmonic oscillator with a spin orbital interaction.

A study of the algebras of integrals of motion and of the corresponding solutions of the Pauli-Schr\"odinger equation is in progress, as is a systematic search for integrals that are two index tensors or pseudotensors.

\begin{table}
\caption{Superintegrable potentials and their integrals of motion}
  \begin{tabular}{ c c c c c c }
    \noalign{\bigskip}
    No & $V_1$ & $V_0$ & Pseudoscalars & Vectors & Axial Vectors \\ \hline \noalign{\bigskip}
    1 & $\frac{\hbar}{r^2}$ & $V_0(r)$ & - & - & (\ref{Saxial})  \\\noalign{\bigskip}
     & & $\frac{\hbar^2}{r^2}$ & - & (\ref{Piaxial}) & (\ref{Saxial})  \\\noalign{\bigskip}
    2 & $\frac{\hbar}{2r^2}$ & $V_0(r)$ & (\ref{pscase1int1}),(\ref{pscase1int2}) & (\ref{firstcasesub11finint3}),(\ref{firstcasesub11finint4}) & -  \\\noalign{\bigskip}
     &  & $\frac{3\hbar^2}{8r^2}-\frac{\alpha}{r}$ & (\ref{pscase1int1}),(\ref{pscase1int2}) & (\ref{firstcasesub11finint1})-(\ref{firstcasesub11finint4}) & -   \\\noalign{\bigskip}
    3 & $-\frac{\hbar}{2r^2}$ & $-\frac{\hbar^2}{8r^2}+\alpha r^2$ & (\ref{intofmotforcase2} ) & - & -   \\\noalign{\bigskip}
    4 & $\frac{3\hbar}{2r^2}$ & $\frac{15\hbar^2}{8r^2}+\alpha r^2$ & (\ref{intofmotforcase3} ) & - & -   \\\noalign{\bigskip}
    5 & $-\frac{\hbar}{r^2}$ & $\alpha r^2$ & - & - & (\ref{axialveccase23})   \\\noalign{\bigskip}
    6 & $\frac{2\hbar}{r^2}$ & $\frac{3\hbar^2}{r^2}+\alpha r^2$ & - & - & (\ref{axialveccase30})   \\\noalign{\bigskip}
    7 & $\frac{\alpha}{2r^2}+\beta$ & $\frac{\alpha(\alpha+2\hbar)+4\beta r^4}{8r^2}$ & - & (\ref{finintofmotionveccase1a}),(\ref{finintofmotionveccase1}) & -   \\\noalign{\bigskip}
    8 & $\frac{\hbar}{2r^2}\left( 1+ \frac{\epsilon}{\sqrt{1+\beta r^2}}\right)$ & $\frac{\hbar^2}{2r^2}\left( 1+ \frac{\epsilon}{\sqrt{1+\beta r^2}}\right)$ & (\ref{case4intofmotfromoldart}), (\ref{case4intofmot}) & (\ref{revisedadditionaliomvector}) & -   \\\noalign{\bigskip}
    9 & $\frac{\hbar}{2r^2}\left( 1+ \frac{2 \epsilon}{\sqrt{1+\beta r^2}}\right)$ & (\ref{case5f2}) & (\ref{case5intofmot}) & - & -   \\\noalign{\bigskip}
    10 & $\frac{\epsilon\hbar}{r^2\sqrt{1+\beta r^2}}$ & (\ref{axialcase4V0sol}) & - & - & (\ref{newaxialcase4intofmot3})   \\\noalign{\bigskip}
    11 & $\frac{\hbar}{r^2}\left( 1+ \frac{\epsilon}{\sqrt{1+\beta r^2}}\right)$ & (\ref{axialcase5V0sol}) & - & - & (\ref{newaxialcase5intofmot3})
\end{tabular}
\end{table}

\section{Acknowledgments}
P.W. thanks Professor A.G.Nikitin for an interesting discussion about superintegrability in systems with spin. The research of P.W. was partly supported by a research grant from NSERC of Canada. The work of \.{I}.Y. is partially supported by the Scientific and Technological Research Council of Turkey (T\"{U}B\.{I}TAK).

%%%%%%%%%%%%%%%%%%%%%%%%%%%%%%%%%%%%%%%%%%%%%%%%%%%%%%%%%%%%%%%%%%%%%%%%%%%%%%%%%%%%%%%%%%%%%%%%%%%%%%%%%%%%%%%%%%%%%%%%%%%%%%%%%%%%%%%%%%%%%%%%%%%%%%%%%%%%%%%%%
%\newpage
%\makeatletter
%\renewcommand{\@biblabel}[1]{$^{#1}$}
%\makeatother

%%%%%%%%%%%%%%%%%%%%%%%%%%%%%%%%%%%%%%%%%%%%%%%%%%%%%%%%%%%%%%%%%%%%%%%%%%%%%%%%%%%%%%%%%%%%%%%%%%%%%%%%%%%%%%%%%%%%%%%%%%%%%%%%%%%%%%%%%%%%%%%%%%%%%%%%%%%%%%%%%%%

\end{document}